\documentclass[manuscript]{aastex}

\usepackage{pdflscape, lscape, psfig, epsf, graphicx}

\slugcomment{Accepted for Publication in the Astrophysical Journal}

\shorttitle{A Super-DLA System}
\shortauthors{Kulkarni et al.}

\begin{document}

\title{A Super-Damped Lyman-alpha QSO Absorber at $z=2.2$ {\footnote {Based on observations collected during program ESO 385.A-0778 at the European Southern Observatory with UVES on the 8.2 m KUEYUN telescope operated at the Paranal Observatory, Chile.}}}

\author{Varsha P. Kulkarni \altaffilmark{}}
\affil{Department of Physics and Astronomy, University of South Carolina,
    Columbia, SC 29208}
\email{kulkarni@sc.edu}

\author{Joseph Meiring \altaffilmark{}}
\affil{Department of Astronomy, University of Massachusetts, 
    Amherst, MA 01003}
    
 \author{Debopam Som \altaffilmark{}}
\affil{Department of Physics and Astronomy, University of South Carolina,
    Columbia, SC 29208}
    
\author{Celine P\'eroux \altaffilmark{}}
\affil{Laboratoire d'Astrophysique de Marseille, OAMP, Universite Aix-Marseille,  
13388 Marseille cedex 13, France}

\author{Donald G. York \altaffilmark{2}}
\affil{Dept. of Astronomy \& Astrophysics, University of Chicago, Chicago, IL 60637}

\author{Pushpa Khare}
\affil{Inter-University Center for Astronomy \& Astrophysics, Pune 411007, India}

\and

\author{James T. Lauroesch}
\affil{Dept. of Physics \& Astronomy, University of Louisville, Louisville, KY 40292}

\begin{abstract}
We report the discovery of a ``super-damped'' Lyman-alpha absorber at $z_{abs}=2.2068$ toward QSO Q1135-0010 in the Sloan Digital Sky Survey, and follow-up VLT UVES spectroscopy. Voigt profile fit to the DLA line indicates log $N_{\rm H I} = 22.05 \pm 0.1$. This is the second QSO DLA discovered to date with such high $N_{\rm H I}$. We derive element abundances  [Si/H] = $-1.10 \pm 0.10$, [Zn/H] = $-1.06 \pm 0.10$, 
[Cr/H] = $-1.55 \pm 0.10$, [Ni/H] = $-1.60 \pm 0.10$, [Fe/H] = $-1.76 \pm 0.10$, [Ti/H] = $-1.69 \pm 0.11$, [P/H] = $-0.93 \pm 0.23$, and [Cu/H] = $-0.75 \pm 0.14$. 
Our data indicate detection of Ly-$\alpha$ emission in the DLA trough, implying a star formation rate of $\sim$10 
$M_{\odot}$ yr$^{-1}$ in the absence of dust attenuation. C~II$^{*} \, \lambda 1336$ absorption is also detected, suggesting SFR surface density  $-2 < {\rm log} \,  \dot{\psi_{*}} < 0$ $M_{\odot}$ yr$^{-1}$ kpc$^{-2}$. We estimate electron density in the range $3.5 \times 10^{-4}$ to 24.7 cm$^{-3}$ from C II$^{*}$/C~II, and  $\sim$0.5-0.9 cm$^{-3}$ from Si II$^{*}$/Si II. Overall, this is a robustly star-forming, moderately enriched absorber, but with relatively low dust depletion. Fitting of the SDSS spectrum yields low reddening for Milky Way, LMC, or SMC extinction curves. No CO absorption is detected, and C I absorption is weak. 
The low dust and molecular content, reminiscent of some SMC sight-lines, may result from the lower metallicity, and a stronger radiation field (due to higher SFR). Finally, we compare this absorber with other QSO and GRB DLAs. 

\end{abstract}


\section{Introduction} 
A  flux-independent probe  of galaxy evolution is provided by absorption lines in spectra of quasi-stellar objects (QSOs) superposed by 
foreground galaxies along the sightlines,  particularly 
the damped Lyman-$\alpha$ (DLA; neutral hydrogen column densities $N_{\rm HI} \ge 
2 \times 10^{20}$ cm$^{-2}$) and sub-DLA 
($10^{19} < N_{\rm HI} < 2 \times 10^{20}$ cm$^{-2}$) absorbers. DLAs and sub-DLAs are the primary neutral gas reservoir available for star 
formation 
(e.g.,  {\em Storrie-Lombardi \& Wolfe 2000; P\'eroux et al. 2005; Prochaska et al. 2005}). Furthermore, they offer among the most accurate element abundance measurements in distant galaxies. The neutral hydrogen and metal content of DLAs/sub-DLAs provide unique indicators of the star formation history, regardless of galaxy 
redshift or morphology.

The calculation of the co-moving density of neutral hydrogen  ($\Omega_{HI}$) directly relies on the integration of the QSO absorber $N_{\rm H I}$ column density distribution, $f(N_{\rm H I})$. The slope of $f(N_{\rm H I})$ at large $N_{\rm H I}$ is typically $\alpha = -3.5$, implying that systems with
large column density are very rare. Noterdaeme et al. (2009) reported a definite
detection from UVES high spectral resolution data of the first DLA with log $N_{\rm H I} = 22.0  \pm 0.1$, at $z_{abs} = 3.286$ 
towards SDSS J081634+144612. This was the highest H I column density DLA reported so far along
a QSO line of sight. Such a large H I column density is
similar to that of DLAs detected at the redshifts of some Gamma-Ray Bursts (GRBs; e.g. Watson et al. 2006; Ledoux et al., 2009). Existence of such strong QSO DLAs puts a strong constraint on the shape of the H I column density distribution. Indeed, from the fit of the H I column density frequency distribution, Noterdaeme et al. (2009) argue that no more than one system like this is expected in the whole SDSS survey. 

Recently, we serendipitously discovered another such high $N_{\rm H I}$ DLA at $z_{abs} = 2.2068$ in the spectrum of QSO SDSS J113520.39-001053.5 with $z_{em} = 2.907$ (see Fig. 1). The QSO is not a broad-absorption-line QSO. The SDSS spectrum indicates that the DLA has log $N_{\rm H I}  \sim 21.9$. This system is probably missing from the sample of Noterdaeme et al. (2009) because it is outside their automated statistical search which has a fixed minimum redshift per spectrum corresponding to a threshold in $S/N$. The damped Lyman-alpha line is in fact not completely covered in the SDSS, and the system falls below the low-redshift edge of the 
redshift sensitivity function of both Noterdaeme et al. (2009) and Prochaska \&  Wolfe (2009). 

Several strong heavy element lines are seen in the SDSS spectrum which suggest a relatively metal-rich absorber. This DLA is thus an ideal place to search for rare elements and molecules. Of course, to obtain accurate  abundances, higher resolution spectra are essential. Here we report follow-up VLT UVES spectroscopy of this unique sightline.

\section{Observations and Data Reduction}

SDSS J113520.39-001053.5 was observed in queue mode on April 13, 2010 with VLT-U2  using UVES (Dekker et al. 2000). Observing time of about 16,200 s was awarded for this program, but integrations of only 4,760 s were  actually obtained due to scheduling constraints.  The average seeing was 0.99 $\arcsec$. Two settings centered at 
$\lambda_{c}=390$ nm and $\lambda_{c}=564$ nm 
were used to cover observed wavelengths in the range of about 3290-6648 {\AA} 
(with some gaps in the ranges of 4518-4625, 5600-5674). This corresponds to a DLA rest-frame coverage of 1026-1409, 1442-1746 and 1769- 2073{\AA}, giving access to  key lines of many metal ions.  A 1$\arcsec$ slit was used, with $2 \times 2$ binning. 

The data were analyzed with the UVES pipeline, IDL and IRAF.  The red side was re-processed with the latest reflex pipeline. The blue side was binned down to $R \sim 6000$ assuming a 3 pixel resolution element, to improve the S/N. This resolution is adequate for fitting the damped Lyman-alpha line. The red side data were binned by a factor of 2, as these spectra, taken with a 1$\arcsec$ slit and $1 \times 1$ binning, were well oversampled. The binned spectra have a resolution $R \sim 45000$  per 3 pixel resolution element.The continuum levels were determined using cubic spline fits to $\sim 20$ {\AA}-wide individual sections of spectra. The absorption lines were fitted with Voigt profiles using the program FITS6P that we have used extensively in previous studies of interstellar and intergalactic matter. FITS6P (Welty et al. 1991)  evolved from the code used by Vidal-Madjar et al. (1977), and minimizes the $\chi^{2}$ between the data 
and the theoretical Voigt profiles convolved with the instrumental profile. 

\section{Results}

Fig. 2 shows the damped Lyman-alpha absorption profile covered in our UVES blue spectrum.  Fig. 3 and 4 show velocity plots of key absorption lines of various ions. Voigt profile fitting gives H I column density of log $N_{\rm H I} = 22.05 \pm 0.10$. The solid red and dashed blue curves overlaid on the data in the bottom panel of Fig. 2 show the Voigt profile fits corresponding to log $N_{\rm H I} = 22.05 \pm 0.10 $. The upward points in the region 4030-4040 {\AA} are near the location of the Ly-$\beta$ and O VI emission lines of the QSO, and also comparable to the noise level (the latter is shown in the top panel of Fig. 2 as a dashed black curve). It is clear that the absorber has a very high $N_{\rm H I}$. For comparison, we have also shown in dot-dashed orange curves the Voigt profiles for 
log $N_{\rm H I} = 21.85$ and 22.25, which are clearly too extreme. Indeed, log $N_{\rm H I} = 21.85$ is too low to explain the strong core of the line even in the region $\sim 3920-3960$ {\AA} which appears relatively free from blends with other lines. We adopt log $N_{\rm H I} = 22.05 \pm 0.1$ as our best estimate of the H I column density. We use this $N_{\rm H I}$ estimate in deriving the element abundances below, since the contribution of ionized gas is not expected to be large at such a high H I column density, and since the H$_{2}$ contribution is hard to determine (see section 4.1.1).  We note that the DLA is not a superposition of two smaller column density clouds, which would otherwise have been reflected in the metal lines as well. 

\subsection{Absolute Abundances}
The spectra show lines of many ions--low ions such as C I, C II, C IV, O I,  Al II, Al III, Si II,  Si IV,  P II, Ti II, Cr II, Fe II, Ni II, Cu II, and Zn II.  The multi-component Voigt profile fits to the metal lines are shown overlaid on the data in Fig. 3 and 4. Tables 1 and 2 list the column densities derived for individual components. For each component, we have listed the mean velocity relative to $z=2.2068$, the Doppler $b$ parameter, and the column density. The element abundances summed over all the velocity components were calculated with reference to solar levels from Lodders (2003), and are listed in Table 3.  We infer [Zn/H] = $-1.06 \pm 0.10$, [Fe/H] = -1.76 $\pm 0.10$, [Cr/H] = $-1.55 \pm 0.10$, [Ni/H] = $-1.60 \pm 0.10$, [Cu/H] = $-0.75 \pm 0.14$, [P/H] = $-0.93 \pm 0.23$, [Ti/H] = $-1.69 \pm 0.11$ and 
[Si/H] = $-1.10 \pm 0.10$. Overall, this is a moderately enriched absorber. Some of the column density measurements have substantial uncertainties, especially those for P and Cu. We note, however, that, to the best of our knowledge, this is the first detection of Cu in a DLA. 

\subsection{Relative Abundances} 

To examine the nature of dust depletion, we calculated the relative abundances [Fe/Zn], 
[Cr/Zn], [Ni/Zn], [Cu/Zn], [Cu/Fe], [Fe/Si], [Ti/Si], and [P/Si] in the individual velocity components, 
and compared them to the interstellar warm cloud depletion values for the Milky Way, based on a combination of  Welty et al. (1999), Cartledge et al. (2006), and Jenkins (2009). We find that there is considerable variation in the relative abundances from component to component, indicating variation in dust content. Surprisingly, Ti is the most depleted element in only two of the seven velocity components that show Ti detections, but is less depleted than Fe, Ni, and/or Cr in the remaining components, 
with Ni and Cr showing smaller abundances than Fe relative to Zn by 0.2-0.5 dex in all but one component. It is also interesting to note that the abundance of Ni relative to Cr is inverted in the component at $v = 62.5$ km s$^{-1}$. The relative differences between Ti and Fe, Cr, Ni are reminiscent of some velocity components in the SMC interstellar sightline toward Sk 155 (e.g. the [Ti/Zn] variations in Welty \& Crowther 2010). For most abundance ratios such as [Fe/Zn], [Cr/Zn], [Ni/Zn], [Fe/Si], the values observed in the DLA under study are higher (indicating less severe depletions)  than in the Milky Way warm diffuse clouds, by differing amounts in different velocity components. However, [Cu/Fe], [Ti/Si], and [P/Si] are below the warm diffuse cloud levels for some velocity components. This may suggest different dust grain compositions in the super-DLA. On the other hand, the low  
[Cu/Fe] and [P/Si] values in some components may be intrinsic nucleosynthetic differences, reflecting the odd-even effect.

Fig. 5 shows the relative abundances as a function of the Zn II column density in the velocity components with detectable Zn. The Pearson product-moment correlation coefficient for [Ni/Zn] vs. log $N_{\rm Zn \,II}$ is -0.961, which suggests a fairly strong anti-correlation (the 1-tailed probability of the correlation happening by pure chance being 0.0003).  For [Fe/Zn] vs. log $N_{\rm Zn \,II}$, the Pearson product-moment correlation coefficient is -0.455, which implies no significant correlation (1-tailed probability = 0.152). For [Ti/Si], the coefficient is 
-0.823, with a 1-tailed probability of 0.0221 of this correlation arising purely by chance. Thus, the relative abundances of some elements 
are consistent with being correlated with the amount of metals (traced by the Zn II column density). While this trend is far from robust (since it pertains to dust in a small number of velocity components of one absorber), it is reminiscent of the total [Cr/Zn] vs. total metallicity correlation reported for DLAs and sub-DLAs (e.g., Ledoux et al. 2003, Meiring et al. 2009a,b and references therein).  We note, however, that such correlations cannot always be explained by dust, because similar correlations are also observed when the relative abundance does not indicate dust depletion, e.g., for [S/Zn] vs log $N_{\rm Zn \,II}$ in both QSO DLAs and GRB DLAs (e.g., Savaglio et al. 2012). 

Fig. 6 shows the total abundances (summed over all velocity components) of Cu, P, Si, Fe, Ni, and Ti relative to Zn, plotted as a function of the condensation temperature of the element (the latter adopted from Lodders 2003). For each element, the black circles denote the data points for the DLA toward Q1135-0100. The red squares and blue diamonds denote the relative abundances for warm and cold diffuse interstellar clouds in the Milky Way. Elements with higher condensation temperatures normally show higher depletions in the Milky Way clouds, with 
warm clouds showing smaller degree of depletions than cold clouds. The relative abundances [P/Zn] and [Si/Zn] in the DLA toward Q1135-0100 are close to zero, at levels expected for warm Milky Way clouds. The relative abundances [Cr/Zn], [Fe/Zn], [Ni/Zn], and [Ti/Zn] are higher, as expected for the Milky Way clouds, but are less severe than even those in the warm clouds. For Cu, there is a large excess, with [Cu/Zn] being super-solar at about the 2 $\sigma$ level. This is surprising, since both depletion and intrinsic nucleosynthesis (odd-even effect) would predict [Cu/Zn]  $< 0$. 

\subsection{Other Ions}  

One signature of a very gas-rich DLA is the presence of weak C I lines. In the DLA toward Q1135-0010, C I $ \lambda$ 1656 is detected, although in only the velocity component at -39 km s$^{-1}$. Voigt profile fitting gives $N( \rm C I) = (8.89 \pm 1.51) \times 10^{12}$. The latter is consistent with the non-detection of C I $ \lambda$ 1560. The observed ratio of C I/ H I is low for the high H~I column density, compared to the Milky Way, 
and may be a result of the lower metallicity and a stronger radiation field.

A few higher ionization species are also detected in this absorption system. Table 4 lists the column densities for these estimated using Voigt profile fitting. Fig. 7 shows a velocity plot of these ions. 

\subsection{Lyman-$\alpha$ Emission}

Despite the noise in our spectrum, there is a hint of Ly-$\alpha$ emission near the center of the DLA absorption trough. 
To better assess the strength of this feature, we resampled the data by varying factors and found the feature present in all cases. Fig. 8a shows the UVES data resampled by factors of 30 and 70, along with the SDSS data. Fig. 8b shows the UVES data resampled by a factor of about 70 (to a resolution of 1 {\AA}), superposed with the same data box-car 
smoothed by factors of 5, 11, and 21. It is clear that the feature does not go away. In the original SDSS spectrum, 
the integrated flux in the feature is  $(2.87 \pm 0.63) \times 10^{-16}$  erg s$^{-1}$ cm$^{-2}$, where the flux uncertainty is estimated as the average photon noise uncertainty outside the feature at the bottom of the DLA trough, multiplied by the square root of the number of pixels covered by 
the feature. In the UVES spectrum resampled by factors of 30, 50, and 70 (i.e., to resolutions of 0.42, 0.70, and 0.98 {\AA} respectively),  the integrated counts in the feature are $29.1 \pm 10.9$, $46.3 \pm 12.2$, and $77.0 \pm 16.0$, respectively, indicating detections at levels of $\approx 2.7, 3.8,$ and $4.6 \sigma$ levels.  
[The resultant estimates of the integrated flux in the Ly-$\alpha$ emission line would be $(1.15 \pm 0.45) \times 10^{-16}$, $(1.15 \pm 0.30) \times 10^{-16}$, and $(1.11 \pm 0.20) \times 10^{-16}$ erg s$^{-1}$ cm$^{-2}$ respectively, if the flux calibration of the UVES spectrum were used. The difference between the SDSS and VLT line flux values arises due to differences in the detailed structure of the observed Ly-$\alpha$ emission features in the SDSS and VLT data, and differences in the absolute flux calibration of the SDSS and VLT data.]

We use the SDSS spectrum to estimate the absolute flux in the Ly-$\alpha$ emission feature, since the SDSS absolute photometric calibration is believed to be much more accurate. Thus we adopt the integrated Ly-$\alpha$ emission flux to be $ (2.87 \pm 0.63) \times 10^{-16}$ erg s$^{-1}$ cm$^{-2}$. Assuming H$_{0} = 70$ km s$^{-1}$ Mpc$^{-1}$, $\Omega_{m} = 0.3$, and $\Omega_{\Lambda} = 0.7$, the luminosity distance at $z=2.2068$ is 17.522 Gpc. We thus estimate a Ly-$\alpha$ luminosity of  $1.05 \times 10^{43}$ erg s$^{-1}$, i.e. $2.7 \times 10^{9} L_{\odot}$. Assuming case-B recombination and adopting the Kennicutt (1998) relation between the star formation rate (SFR) and H-$\alpha$ luminosity, we estimate SFR = $9.1 \times 10^{-43} \, L_{Ly-\alpha} =  9.6 \pm 2.1 $ M$_{\odot}$ yr$^{-1}$. These estimates assume no correction for dust absorption or resonant scattering by neutral gas. As we discuss in section 4.3, there is not much evidence of a significant amount of dust in this DLA. The inferred Ly-$\alpha$ emission and SFR, if confirmed with higher S/N spectra in the future, is much stronger than in most other QSO absorbers (e.g., Bouch\'e et al. 2007; Fynbo et al. 2011;  Kulkarni et al. 2000, 2006; Rahmani et al. 2010; P\'eroux et al. 2011a, 2011b, and references therein). Indeed, as noted by Wolfe \& Chen (2006), the star-formation efficiency for high-redshift DLAs is low. We discuss the star formation in the DLA toward Q1135-0100  further in section 4.4 using C II$^{*}$ absorption.

\section{Discussion}

\subsection{Comparison with Other High-$N_{\rm H I}$ Absorbers}

We now compare the super-DLA toward Q1135-0100 with other DLAs observed toward QSOs and GRBs. The comparison QSO DLA sample is from the compilation used in Kulkarni et al. (2010), while the GRB DLA sample is based on Prochaska et al. (2007, 2009), 
Ledoux et al. (2009), Rau et al. (2010), and Savaglio et al. (2012).  For ease of comparison, we have only used systems with firm measurements of Zn (in most cases) or S (in a few cases with no Zn detections), excluding upper or lower limits. Both Zn and S are almost undepleted in the warm Milky Way interstellar clouds, and can thus serve as 
good indicators of the total metallicity, without the need to correct for dust depletion. 

\subsubsection{Metallicity , H I Content, and Molecular Gas}

The possibility that some high-redshift QSOs may be obscured by dust in foreground DLAs was suggested by Fall \& Pei (1993). Boiss\'e et al. (1998) pointed out that there is a deficit of DLAs with high $N_{\rm H I}$ and high metallicity. Since the metal lines of such 
systems should be easy to measure, Boiss\'e et al. postulated that the deficit is a selection effect caused by dust obscuration: systems with large H I column densities and high metallicities would have larger metal column densities and hence larger dust column densities. Such absorbers could therefore obscure the background QSOs enough that they would tend to be underrepresented in spectroscopic studies that often tend to favor a brighter QSO over a fainter QSO. 
Boiss\'e et al. suggested an empirical obscuration threshold of log $N_{\rm Zn II}$ = 13.15. Over the past few years, a  few DLAs above this threshold have been discovered toward QSOs and GRBs (e.g., Meiring et al. 2006; Peroux et al. 2008), thereby casting doubt on the empirically inferred existence of the obscuration threshold. Several of the systems detected beyond the threshold, however, lie close to the threshold. 

Fig. 9 shows a plot of the metallicity vs. H I column density for QSO DLAs and GRB DLAs.  The QSO DLAs are from Kulkarni et al. (2010) and references therein. The GRB DLAs are from Vreeswijk et al. (2004), Prochaska et al. (2007), Ledoux et al. (2009), Savaglio et al. (2012), and references therein. The solid orange line denotes the obscuration threshold suggested by Boiss\'e et al. (1998). 
The DLA toward Q1135-0100 is significantly beyond the ``obscuration threshold''; in fact, it is the ``farthest" QSO DLA from the obscuration threshold known so far. In terms of the ``distance" from the obscuration threshold line, the DLA toward Q1135-0100 is in fact, comparable, to the DLA toward GRB 000926.  The DLA toward GRB080607 is believed to be even further beyond in the ``zone of avoidance", although its metallicity is highly uncertain owing to line saturation (Prochaska et al. 2009). 

A possible explanation of the low $N_{\rm HI}$ in QSO-DLAs was suggested by Pontzen et al. (2010): QSO-DLAs are in random sight lines, likely in the external part of galaxies where most of the volume is and where gas density and metallicity are lower, whereas GRB-DLAs are in star-forming regions closes to the galaxy center, where gas column densities are on average larger. However, this does not explain why the metallicities of most QSO DLAs are lower than those in QSO sub-DLAs (e.g., Kulkarni et al. 2010 and references therein), which should be tracing gas even further out. 

Another explanation of the deficit of DLAs with high $N_{\rm H I}$ and high metallicity was suggested by Schaye et al. (2001) and Krumholz et al. (2009a), in terms of the formation of molecules in the cold phase of a 2-phase interstellar medium in pressure balance.  Their idea is that 
above some $N_{\rm H I}$ threshold that decreases with increasing metallicity, the hydrogen becomes predominantly molecular, 
and hence undetectable in observations of the Ly-$\alpha$ absorption line. A fixed optical depth of material can absorb the entire 
photon flux in the Lyman/ Werner region, so any additional gas is molecular. A higher metallicity implies a higher dust-to-gas ratio, 
which results in an increased H$_{2}$ formation rate. This, because of the increased cooling via metals and H$_{2}$, result in a lower temperature. Both the higher H$_{2}$ formation rate and the lower temperature lead to a decrease in the maximum $N_{\rm H I}$.

The remaining lines in Fig. 9 show curves calculated by Krumholz et al. (2009a) for ``covering'' fractions  $c_{\rm {H_{2}}}$ = 0.01, 0.05, 0.5, and 1.0 for the cross-section of the spherical molecular core of the cloud (surrounded by a shell of atomic gas).  These values of $c_{\rm H_{2}}$ correspond to molecular mass fractions $f_{\rm H_{2}} = M_{\rm H_{2}}/M_{\rm total} = [1 -  \{(1 - c_{\rm H_{2}}^{-1.5})/ \phi_{mol} \}]^{-1} = 0.010$, 0.102, 0.845, and 1.0, respectively. [The quantity $\phi_{mol}$ is the ratio of molecular gas density to atomic gas density, and is about 10 (Krumholz, McKee, \& Tumlinson 2009b)]. In this simple picture, the DLA toward Q1135-0100 may seem to be consistent with an H$_{2}$ core ``covering" fraction of $\gtrsim 0.05$ (although values as small as $\sim 0.03$ and as high as $\sim 0.5$ cannot be ruled out within the uncertainties on the metallicity and H I column density). 
Thus the core ``covering" fraction could be higher than the values inferred to be $< 0.06$ by Krumholz et al. (2009a) for most other QSO or GRB DLAs (except that toward GRB080607; Prochaska et al. 2009).  

In this scenario, the molecular mass fraction $f_{\rm H_{2}}$ for the DLA toward Q1135-0010 could be as low as 0.05 or as high as 0.85--much higher than that observed for other QSO DLAs, in either case. This DLA would therefore be a good candidate to search for molecules. Unfortunately,  the Lyman and Werner bands of H$_{2}$ in the DLA lie in the very noisy part of the spectrum affected by the Lyman limit break from the absorber at $z=2.92$, and are thus unobservable. No radio and sub-mm observations exist either, so no information exists on the molecules from the DLA. We attempted to search for UV absorption features of CO AX 0-0, 1-0 , 2-0, 3-0, and 4-0 near 1544, 1510, 1478, 1447, 1419 {\AA} respectively, but no significant features were detected with a consistent set of velocity components. From the non-detection of the strongest of these features (CO AX 0-0), we estimate the 3 $\sigma$ upper limit to the CO column density as log $N_{\rm CO} < 13.80$, including the effect of both photon noise uncertainty and continuum determination uncertainty. This CO column density limit is comparable to the $N_{\rm CO}$ detected in a QSO DLA by Srianand et al. (2008b), but much smaller than that in the log $N_{\rm HI} = 22.7$ DLA toward GRB 080607 (Prochaska et al. 2009). 

The non-detection of CO may seem surprising because of the high molecular mass fraction suggested by Fig. 9, and because a substantial H$_{2}$ fraction may be expected at such a high $N_{\rm H I}$  
in the interstellar medium of the Milky Way. However, it is consistent with the low observed C I column density $N_{\rm  C I} \approx 10^{13} $ cm$^{-2}$. This value of $N_{\rm C I}$ is close to the regime where H$_{2}$ becomes self-shielded, so  $N_{\rm H_{2}} $ may range anywhere from $\sim 10^{14}$ to $\sim 10^{19}$ cm$^{-2}$ (see, for example, Fig. 22 and 27 in Welty \& Hobbs 2001). Such a range in H$_{2}$ column density is entirely consistent with the inferred upper limit on CO (log $N_{\rm CO} < 13.8$). We also note that the Krumholz et al. (2009a) scenario has some caveats (e.g., the assumption of 2-phase equilibrium and the exclusion of warm H I) which may not apply in this case (for example since there is clearly evidence of warm gas in this DLA).  Such cases of low $N_{\rm H_{2}}$ at high $N_{\rm H I}$ are not common, but not unprecedented. For example, Welty, Xue, \& Wong (2012) report sightlines 
toward $\theta ^{1}$ Ori C in the Orion Trapezium region and Sk9 in the southwestern part of the main SMC bar that also show low $H_{2}$ at high $N_{\rm H I}$, presumably because of strong radiation fields. Indeed, SMC sightlines with log $N_{\rm H I} \sim 22$ show a wide range of the number fraction of $H_{2}$ from $\sim 10^{-6}$ to $10^{-1}$ (e.g., Tumlinson et al. 2002). Thus we conclude the molecular fraction is low in the DLA toward Q1135-0010 because of the low metallicity, low dust content (see sec.  4.2, 4.3) and/or strong radiation field. It will be interesting to determine the molecular content of this absorber more accurately with higher S/N spectra in the future. 
 
\subsubsection{Dust Depletion vs. H I Content}

Watson et al. (2006) reported a DLA with log $N_{\rm HI} = 22.6$ toward the afterglow of GRB 050401. They reported 
[Si/H] = -1.7, [Zn/H]=-1.0, [Fe/H] = -2.1, and [Cr/H] = -1.7. Compared to this, the DLA toward Q1135-0010 has about the same metallicity, but a smaller amount of depletion. In particular, Si is essentially undepleted relative to Zn ([Si/Zn]= $-0.04 \pm 0.04$), compared to [Si/Zn] =-0.7 in the GRB DLA of Watson et al. (2006). We note,however, that the measurements for the DLA toward GRB 050401 have large uncertainties ([Si/Zn] $=-0.7 \pm 0.5$), so the latter could be consistent with the [Si/Zn] for the DLA toward Q1135-0010 within the uncertainty. While [Fe/Zn] and [Cr/Zn] are lower  compared to [Si/Zn] in the DLA toward Q1135-0010, they are not as low as in the DLA toward GRB 050401. For the afterglows of GRB 050730 ($z_{abs} = 3.969$), GRB 071031 ($z_{abs} = 2.692$), and GRB 080413A ($z_{abs} = 2.435$), 
Ledoux et al. (2009) report log $N_{\rm H I}$ = 22.10, 22.15, and 21.85, respectively. These latter absorbers show 
metallicities [X/H] of $-2.18 \pm 0.11$, $-1.73 \pm 0.05$, and $-1.60 \pm 0.16$ (where X = S, Zn, and Zn, respectively). 
The corresponding relative abundances are [Fe/X] = $0.06 \pm 0.06$, $-0.04 \pm 0.02$, and $-0.13 \pm 0.07$, all of which are 
surprisingly low. Indeed, Ledoux et al. (2009) show that the majority of the GRB DLAs in their sample lie at the high-$N_{\rm H I}$ , low [X/Fe] end in the plot of $N_{\rm H I}$ vs. [X/Fe]. Fig. 10 shows [Fe/X] vs. H I column density (where X = Zn in most cases, S in a few) for QSO DLAs compiled in Meiring et al. (2009b), along with GRB DLAs from Prochaska et al. (2007) and Ledoux et al. (2009), along with the DLA toward Q1135-0010. It is clear that the 
DLA toward Q1135-0100 lies in the less-populated  central right part of this plot (high H I column density, moderately large [Fe/X]). While the relative abundance [Fe/Zn] observed for this DLA is not the largest, it is in the upper half of the range of [Fe/Zn] observed for QSO DLAs at $z > 1.5$. 

\subsection{QSO Reddening}

The SDSS magnitudes of Q1135-0010 are $u, g, r, i, z = 21.22 \pm 0.08, 19.03 \pm 0.01, 18.47 \pm 0.01, 18.25 \pm 0.01, 18.25 \pm 0.02$, respectively. The $\Delta (g-i)$ color, relative to the median $(g-i)$ for the SDSS composite QSO spectrum at $z_{em} = 2.907$ (adopted from Richards et al. 2001) is 0.48. If most of this apparent reddening arises in the DLA at $z=2.2068$, the implied $E(B-V) = \Delta(g-i) (1+z_{abs})^{-1.2} / 1.506$ would be 0.08, assuming the SMC extinction curve.  There are two other strong absorbers in this sightline at $z=2.696$ and $z=2.924$, but their Ly-alpha lines are much weaker (rest frame equivalent widths of 1.08 and 1.13 {\AA}, implying log $N_{\rm H I} < 19$). Thus these other absorbers are not even sub-DLAs), and are not likely to contribute much compared to the DLA with log $N_{\rm H I} = 22.0$. However, part of the inferred $E(B-V)$ for the DLA toward Q1135-0010 arises from the fact that  the strong DLA line falls within the $g$ band and removes a large section of the QSO continuum, so the true reddening is likely even lower.

The implied amount of low reddening in this DLA is comparable to that in most QSO absorbers at $z \sim 2$. For example, the reddening found by York et al. (2006) in a study of SDSS Mg II absorbers at $1 \le z < 2$ is $< 0.09$, while that found by  for SDSS DLAs at $2.2  \le z < 5.2$ is $-0.0017 \pm 0.0022$ (Frank \& P\'eroux 2010) or $< 0.03$ (Khare et al. 2012). 

\subsection{Extinction Curve Fitting}

We note that the SDSS spectrum shows no evidence of a 2175 {\AA} bump. Thus the dust in this absorber does not appear to resemble that in the Milky Way.  To quantify the reddening, we fitted the SDSS flux-calibrated spectrum with the Milky Way, LMC, LMC super-shell (LMC2), and SMC bar extinction curves, following the procedure described in Kulkarni et al. (2011). The original spectrum of the QSO was taken to be the composite QSO spectrum adopted from Van den Berk et al. (2001). Next, for a given assumed extinction curve, the optical depth at any wavelength in the absorber rest  frame was calculated as $\tau (\lambda) = 0.921 \, A_{\lambda} = 0.921 \, A_{V} \,  \zeta_{\lambda}$, where $A_{\lambda}$ is the extinction at wavelength $\lambda$, $A_{V} = R_{V} \times E(B-V)$ is the extinction in the V band, and $\zeta_{\lambda}$ is the extinction curve, all quantities being in the absorber rest frame. The expected spectrum was then 
calculated as a function of $E(B-V)$ and a flux normalization factor and matched to the dereddened observed spectrum with a least squares approach. The Fitzpatrick-Massa (FM) parameters for the average extinction curves and the $R_{V}$ values  for the Milky Way, Large Magellanic Cloud (LMC), Large Magellanic Cloud supershell (LMC2), and the Small Magellanic Cloud (SMC) bar  from Misselt et al. (1999) and Gordon et al. (2003) were adopted. Emission and absorption lines in the QSO spectrum were excluded while doing the fitting so as not to bias the continuum level. Fig. 11 shows the extinction curve fits to the SDSS spectrum dereddened for Galactic extinction, which is shown here box-car smoothed by a factor of 9 for the purpose of display. The best-fit $E(B-V)$ values for the MW, LMC, LMC2, and SMC bar extinction curves are -0.04, -0.03, 0.03, and 0.02, respectively, with reduced $\chi^{2}$ values of 6.07, 6.70, 6.20, and 5.30, respectively. Thus, among the four extinction curves, SMC bar appears to fit the best. In fact, all four extinction curves suggest low reddening, although none of these extinction curves fits well. (We note that the negative $E(B-V)$ values 
for the MW and LMC extinction curves are not physically meaningful; they just indicate that there is no significant reddening, and in fact, there is no satisfactory fit for those extinction curves. Of course, as in previous studies utilizing similar techniques for extinction curve-fitting of QSO spectra (e.g., Srianand et al. 2008a), 
an individual QSO's spectrum could be different from the composite QSO spectrum template.)

We thus conclude that the DLA toward Q1135-0010 does not show much extinction. This is surprising compared to most  interstellar clouds with high $N_{\rm H I}$ in the Milky Way, which show large extinction. The low extinction observed is consistent, however, with 
the lack of detectable CO lines, and low molecular fraction inferred above. 

\subsection{Cooling Rate and Star Formation Rate Surface Density}

The absorber toward Q1135-0100 offers a unique opportunity to estimate the star formation rate in the absorbing galaxy in two independent ways. As discussed earlier, the tentative detection of the Ly-$\alpha$ emission suggests a star formation rate of  $9.6 \pm 2.1 $ M$_{\odot}$ yr$^{-1}$ (although this detection needs to be confirmed with higher S/N spectra). An additional constraint on the SFR can be obtained from the detection of C II$^{*}$ $\lambda 1336$ absorption, following Wolfe et al. (2003). 

Most of the interstellar cooling is due to the fine structure line emission of [C II] $\lambda$158 $\mu$m, which arises from the $^{2} P_{3/2}$ to $^{2}P_{1/2}$ transition in the ground $2s^{2} \, 2p$ term of C II. The cooling rate can be expressed as $l_{c} = N_{\rm C II^{*}} h \nu_{ul} A_{ul} / N_{\rm H I}$, where $N_{\rm C II^{*}}$ is the column density of the C II ions in the 2P$_{3/2}$ state, while $A_{ul}$ and $h \nu_{ul}$ are the 
coefficient for spontaneous photon decay and energy of the $^{2} P_{3/2}$ to $^{2}P_{1/2}$ transition (e.g., Pottasch et al. 1979).  The UV transition of  C~II$^{*} \, \lambda 1336$ can be used to infer $N_{\rm C II^{*}}$.  
For the absorber toward Q1135-0010, 
the range of $N_{\rm C II^{*}}$ consistent with the data is $1.01 \times 10^{15} < N_{\rm C II^{*}} < 4.66 \times 10^{16}$ cm$^{-2}$. {\footnote {This range of $N_{\rm C II^{*}}$ corresponds to the C II column density range consistent with the C II $\lambda 1334$ line, i.e. $3.02 \times 10^{16}$ cm$^{-2} < 
N_{\rm {C II}} < 2.28 \times 10^{19}$ cm$^{-2}$. Of course, we cannot rule out 
Ly-$\alpha$ forest blends, but it would be remarkable if the blends occurred exactly at the positions of the major velocity components expected in  C II$^{*}$ from the other second ions. While we cannot completely rule out that some of the C II$^{*}$ absorption may be high-velocity C II, no other low ionization species in this DLA show absorption at such high velocities. Therefore this possibility seems remote.}} This implies $ 2.6 \times 10^{-27} <  l_{c} < 1.2 \times 10^{-25}$ erg s$^{-1}$ per H atom. 

In Fig. 12, we plot the cooling rate vs. H I column density for interstellar clouds in the Milky Way, along with the corresponding measurements for DLAs from Wolfe et al. (2003) and the DLA studied here. The ISM measurements are adopted from Lehner et al. (2004) and shown separately for low, low+intermediate, intermediate, and high-velocity clouds in the Milky Way. Most DLA measurements seem consistent with the anti-correlation between $l_{c}$ and $N_{\rm H I}$ seen in the Milky Way interstellar clouds.  The 
uncertainty in the C II$^{*}$ column density for the DLA toward Q1135-0010 mentioned above results in a fairly large uncertainty in the cooling rate. If the cooling rate for the DLA toward Q1135-0100 is close to the upper limit, it would be clearly very distinct from both the ISM clouds in the Milky Way and other DLAs.  

Based on the observed gas-phase abundances [Fe/H] and [Si/H] (the latter almost equal to [Zn/H]), we estimate the dust-to-gas ratio to be about 0.076, using equation 7 of Wolfe et al. (2003). Using this dust-to-gas ratio, the range of cooling rate $l_{c}$ inferred from C~II$^{*}$, and comparing to the predicted cooling rate vs. gas density from calculations similar to Fig. 5 of Wolfe et al. (2003) for  a range of SFR surface density $\dot{\psi_{*}}$ values, one can constrain the cold neutral medium (CNM) density and hence $\dot{\psi_{*}}$. While a detailed calculation of the $l_{c}$ vs gas density relation would give a more exact estimate of $\dot{\psi_{*}}$, given the observational uncertainty in $l_{c}$, here we estimate the approximate range of $\dot{\psi_{*}}$. To do this, we make use of the fact that the observed Fe and Si abundances and lower limit on $l_{c}$ for the $z=2.2068$ DLA toward Q1135-0010 are all within 0.1-0.2 dex of the corresponding quantities for the DLA at the comparable redshift $z=2.04$ toward Q0458-02 from Wolfe et al. (2003). Thus, using Fig. 5b of 
Wolfe et al. (2003), we estimate that $-2 \lesssim $ log $\dot{\psi_{*}} \lesssim 0$ M$_{\odot}$ yr$^{-1}$ kpc$^{-2}$ from the 
limits on $l_{c}$ derived above. 
 It is interesting that the SFR surface density expected from the observed H I column density on the basis of the Kennicutt (1998) relation $\dot{\psi_{*}} = [2.5 \times 10^{-4}] \times [N_{\rm H I} / 1.26 \times 10^{20}]^{1.4}$ would be in the middle of this range (log $\dot{\psi_{*}} =  -0.87$). 

Adopting log $\dot{\psi_{*}} \sim -1$ M$_{\odot} $ yr$^{-1}$ kpc$^{-2}$, and using the SFR $\approx 10$ M$_{\odot}$ yr$^{-1}$ estimated in section 3.4 from the Ly-$\alpha$ emission flux along the line of sight, it may seem that the line-of-sight extent of the DLA absorber should  be $d \sim 10$ kpc in the simple assumption of the star formation being distributed uniformly within the DLA. However, since both estimates of SFR are along the pencil beam sightline toward the QSO (with a projected beam size $\sim$ pc), it appears that the star formation is highly clumpy (otherwise the implied DLA line-of-sight extent would be enormous). 

\subsection{Electron Density}

\subsubsection{C II$^{*}$ Absorption}

The ratio of column densities of C II$^{*}$ and C II can be used to constrain the electron density in the gas, for an assumed gas temperature, because the C II upper level is expected to be populated by collisional excitation, and depopulated by radiative de-excitation in the intervening DLA toward Q1135-0010. Equilibrium between these processes gives  
$N_{\rm C II^{*}} /N_{\rm C II}= n_{e} C_{12}(T)/ A_{21}$, where $A_{21} = 2.29 \times 10^{-6}$ s$^{-1}$ (Nussbaumer \& Storey 1981).  The collision rate coefficient is given by $C_{12}(T) = [8.63 \times 10^{-6} \Omega_{12} / (g_{1} T^{0.5})] \, {\rm exp}(-E_{12}/kT)$ (Wood \& Linsky 1997), where $g_{1} = 2$, $E_{12} = 1.31 \times 10^{-14}$ erg, and the collision strength $\Omega_{12}$ depends on temperature only weakly. At T$\sim 7000$ K, $\Omega_{12} = 2.81$, giving $C_{12} = 1.43 \times 10^{-7}$. 
For the absorber toward Q1135-0010, using the maximum C II and minimum C II$^{*}$ column densities (see section 4.4), we get 
N$_{\rm C II^{*}} / N_{\rm C II} > 4.4 \times 10^{-5}$, which implies the electron density $n_{e} > 7.1 \times 10^{-4}$ cm$^{-3}$. 
A roughly similar  
electron density lower limit is obtained even if 
the gas is much cooler, say $T = 500$ K, since at this temperature 
$\Omega_{12} = 1.82$, giving $C_{12} = 2.90 \times 10^{-7}$, and hence $n_{e} >  3.5 \times 10^{-4}$ cm$^{-3}$. By contrast, if the 
minimum C II and maximum C II$^{*}$ column densities are used, we get the electron density $n_{e} <  24.7$ cm$^{-3}$ for $T = 7000$ K ($n_{e} < 12.2$ cm$^{-3}$ for $T = 500$ K). Of course the upper limit is very high for either temperature considered, and the actual electron density is likely much lower (see section 4.5.2 below).

We note that the maximum C II column density would imply a highly supersolar [C/H], reminiscent of the C-enhanced absorbers 
reported by Cooke et al. (2011) for much more metal-poor absorbers. The true C II (and C II$^{*}$) column densities may be intermediate between the two limiting values discussed above, and the true [C/H] may not be as extreme. Correspondingly the electron density and cooling rates would be intermediate between the limiting values discussed above. 

\subsubsection{Si II$^{*}$ Absorption}

Additional constraints on the electron density can be obtained from the Si II$^{*}$ absorption, which also appears to be present in some of the velocity components for this DLA. To our knowledge, this is the first detection of Si II$^{*}$ in 
an intervening QSO absorber, although Si II$^{*}$ absorption has been reported in GRB afterglows (e.g. Savaglio et al. 2012 and references therein). Fig. 13 shows velocity plots for the Si II$^{*}$ 1264.7, 1265.0, 1309.8, and  1533.4 lines. (The Si II$^{*}$ 1265 panel shows the combined contribution  of the $\lambda$ 1264.7 and $\lambda$ 1265.0 lines as the solid green curve and the contribution of  Si II$^{*}$ $\lambda$ 1265.0 alone as the dashed blue curve.) The green curves show our Voigt profile fits, with the main contributions coming from velocity components near -73, -31, and -19 km s$^{-1}$. These fits yield a total column density (summed over all the velocity components) of log $N_{\rm Si II^{*}} = 13.70^{+0.04}_{-0.05}$, giving N$_{\rm Si II^{*}} / N_{\rm Sii II} = (1.62 \pm 0.18) \times 10^{-3}$.  The Si II$^{*}$/Si II ratio can be used to estimate the electron density assuming equilibrium between collisional excitation and spontaneous radiative de-excitation in this intervening DLA. Using the collisional excitation rate for Si II [given by $C_{12} = 3.32 \times 10^{-7}  \, (T/10,000)^{-0.5} \, exp(-413.4/T)$ cm$^{3}$ s$^{-1}$; e.g., Srianand \& Petitjean (2000)] and the spontaneous radiative de-excitation rate for Si~II$^{*}$ $A_{21} = 2.13 \times 10^{-4}$ s$^{-1}$, 
the corresponding estimate of the electron density is $n_{e} = 0.53 \pm 0.06$ cm$^{-3}$ for $T=500$ K or $0.91 \pm 0.10$ cm$^{-3}$ for $T= 7000$ K.

\subsection{Implications for Co-moving Density of Neutral Hydrogen}

The co-moving density of neutral hydrogen $\Omega_{HI}$ is sensitive to the presence of high $N_{\rm H I}$ systems, since it constitutes an integral of the product of the H I column density and the column density distribution $f(N_{\rm H I})$ which is roughly shaped as a Gamma function or a combination of power laws with negative slopes. We note, along the lines of Noterdaeme et al. (2009), that DLAs might have not been identified in the Sloan Survey because the presence of the DLA trough significantly reduces the QSO flux and hence decreases the corresponding S/N of the spectrum. If many such high N(HI) column densities were missing, this would affect the shape of the column density distribution $f(N_{\rm H I})$ and hence the resulting neutral gas mass, $\Omega_{HI}$ (Prochaska et al. 2005, Noterdaeme et al. 2009). Given the small redshift region ($< 0.1$) covered by the SDSS but not sampled by prior studies (e.g., Prochaska \& Wolfe 2009; Ledoux et al. 2009), and the number density of strong DLAs, we expect very few if any other absorbers similar to the super-DLA among the SDSS DR7 data. Given the rarity of such systems, we anticipate that they will not have a significant 
impact on the estimates of $\Omega_{HI}$. 

On the other hand, such high $N_{\rm HI}$ systems can provide unique insights into the nature of dust, metals, and molecules in the most gas-rich galaxies. It would be interesting to obtain radio / sub-mm observations of Q1135-0100 for followup studies 
of molecular gas in this unique environment. It would also be of great  interest to attempt to image the galaxy responsible for the super-DLA in this sightline. We plan to pursue these directions in the near future. 

{\it Note:} Toward the very end of the refereeing process of this paper, we learned about another analysis of the same object by Noterdaeme et al. (2012), who used VLT X-shooter and Magellan MagE spectroscopy, together with our VLT UVES data. 
The $N_{\rm H I}$ value derived by Noterdaeme et al. (2012) (log N$_{\rm H I} = 22.10 \pm 0.05$) agrees closely with our value of $22.05 \pm 0.10$, as do the abundances of various elements (Si, Cr, Fe, Ni, Zn) that they derived from their X-shooter data and our UVES data. The dust reddening estimated by these authors from the X-shooter spectrum (0.04 for an SMC extinction curve) is also close to our own corresponding estimate (0.02). The Lyman-alpha emission is also confirmed in the X-shooter spectrum, and in fact shows a resolved double-peaked profile, giving a flux within a factor of 2 of our estimate from the SDSS spectrum (and within a factor of 1.25 of the estimate from our VLT UVES spectrum). 
\vskip 1.0in

\acknowledgments

We are grateful to the ESO time allocation committee for the allocation of time for the VLT observing program 385.A-0778 for this project. 
We are also grateful to an anonymous referee and to D. E. Welty for comments that have helped to improve this paper.
V.P.K.  and D. S. acknowledge partial support from NSF grants AST-0908890 and AST-1108830 to the University of South Carolina (PI: Kulkarni). 

{\it Facilities:} \facility{VLT(UVES), SDSS}.

\clearpage

\begin{figure}
\hskip -0.15in
\includegraphics[width=0.8\textwidth, angle=270]{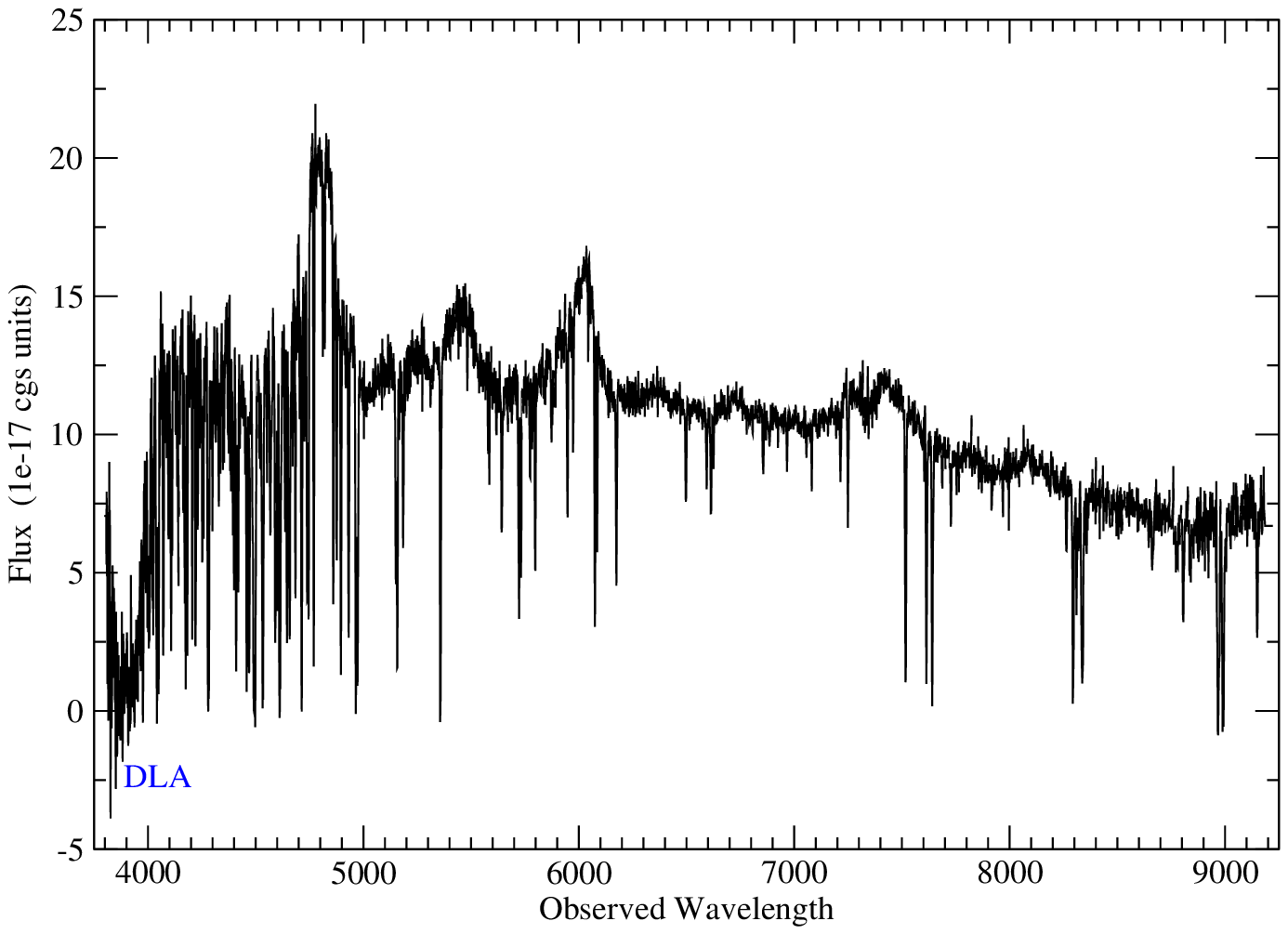}
\hsize=15.0truecm
 \caption{The SDSS spectrum of Q1135-0010 that enabled us to detect the super-DLA. The DLA trough is labeled. }
\end{figure}

\begin{figure}
\hskip -0.15in
\includegraphics[width=0.8\textwidth, angle=270]{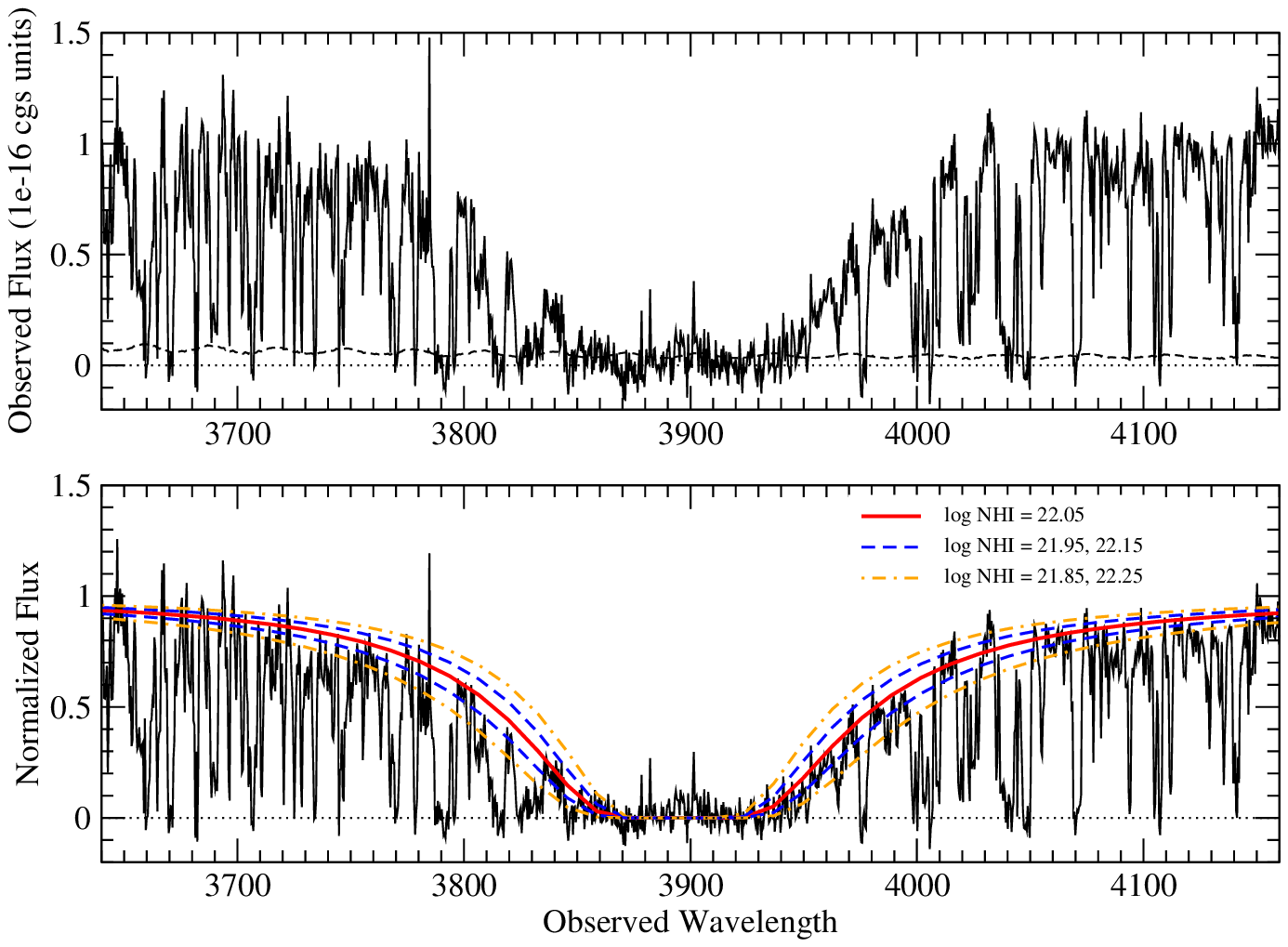}
\hsize=15.0truecm
 \caption{The redshifted damped Lyman-alpha line from the VLT UVES spectrum of Q1135-0010. Top: The solid black curve shows the un-normalized data, resampled to a dispersion of 0.42 {\AA} per pixel, (i.e., 30 times lower resolution than the original data), while the dashed black curve near the bottom indicates the 1 $\sigma$ uncertainty in this resampled spectrum. Bottom: The solid black curve shows the resampled spectrum after continuum normalization. The solid red and dashed blue curves overlaid on the data show the Voigt profile fits corresponding to log $N_{\rm H I} = 22.05 \pm 0.10$, all centered at $z=2.2062$. For comparison, we also show as dot-dashed orange curves  the Voigt profile fits for  log $N_{\rm H I} =   21.85$ and 22.25, which are clearly too extreme.}
\end{figure}

\newpage
\begin{figure}
\hskip -0.15in
\hsize=18.0truecm \epsfysize \hsize \epsfbox{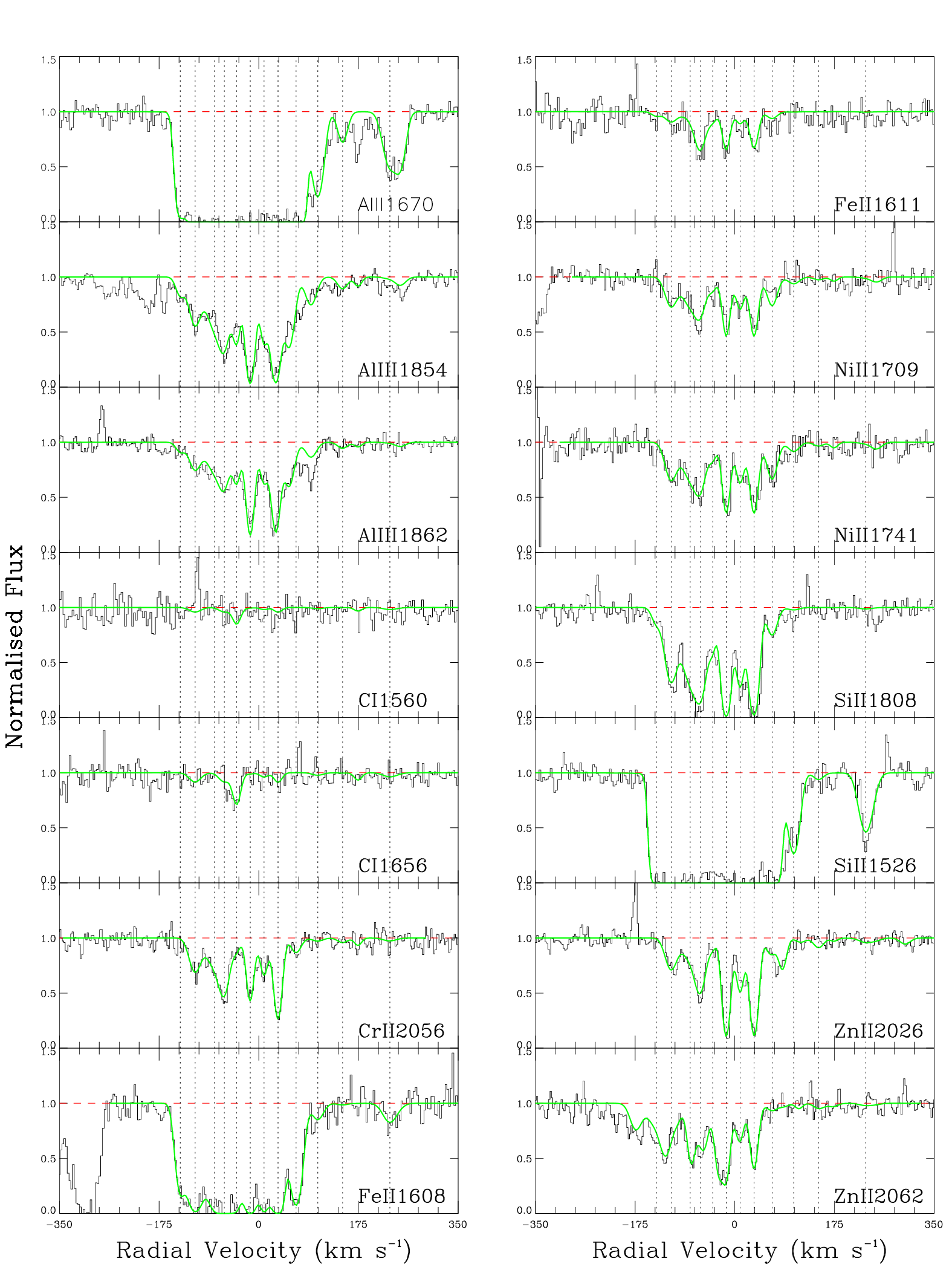}
 \caption{Velocity plots of key metal lines in the $z_{abs} = 2.2068$ absorber toward  Q1135-0010. The green curves indicate the multi-component Voigt profile fits. The dotted vertical lines indicate the positions of the velocity components.} 
\end{figure}

\newpage
\begin{figure}
\hskip -0.15in
\includegraphics[width=0.81\textwidth, angle=90]{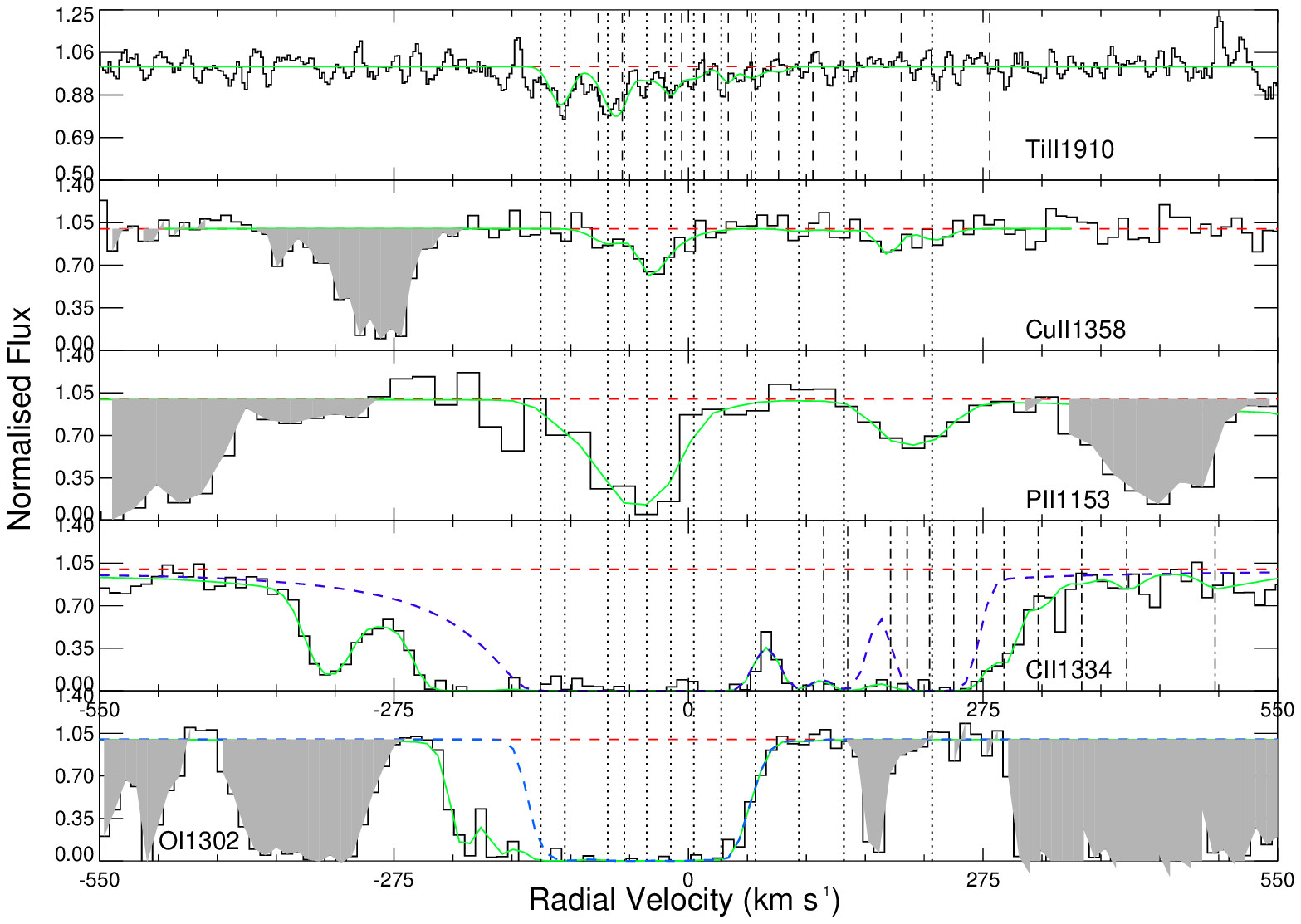}
 \caption{Velocity plots of additional metal lines in the $z_{abs} = 2.2068$ absorber toward  Q1135-0010. The green curves indicate the multi-component Voigt profile fits. The dotted vertical lines indicate the positions of the velocity components. The dotted and dashed vertical lines in the top panel show the positions of the velocity components for Ti II $\lambda \,  \lambda 1910.612, 1910.954$. In the bottom two panels, the dashed blue curves indicate the contributions from C II $\lambda$ 1334 and O I $\lambda$ 1302, the remaining contributions coming from blends with C II$^{*}$ $\lambda 1336$ or Ly-$\alpha$ forest lines. The C II $\lambda 1334$ panel shows the maximum C~II contribution consistent with the data. 
The dot-dashed vertical lines in the C II $\lambda$1334 panel denote the contribution from C~II$^{*}$ $\lambda 1336$. Unrelated lines of systems at other redshifts are blocked out in grey for display purposes. }
\end{figure}

\newpage
\begin{figure}
\includegraphics[width=0.81\textwidth, angle=270]{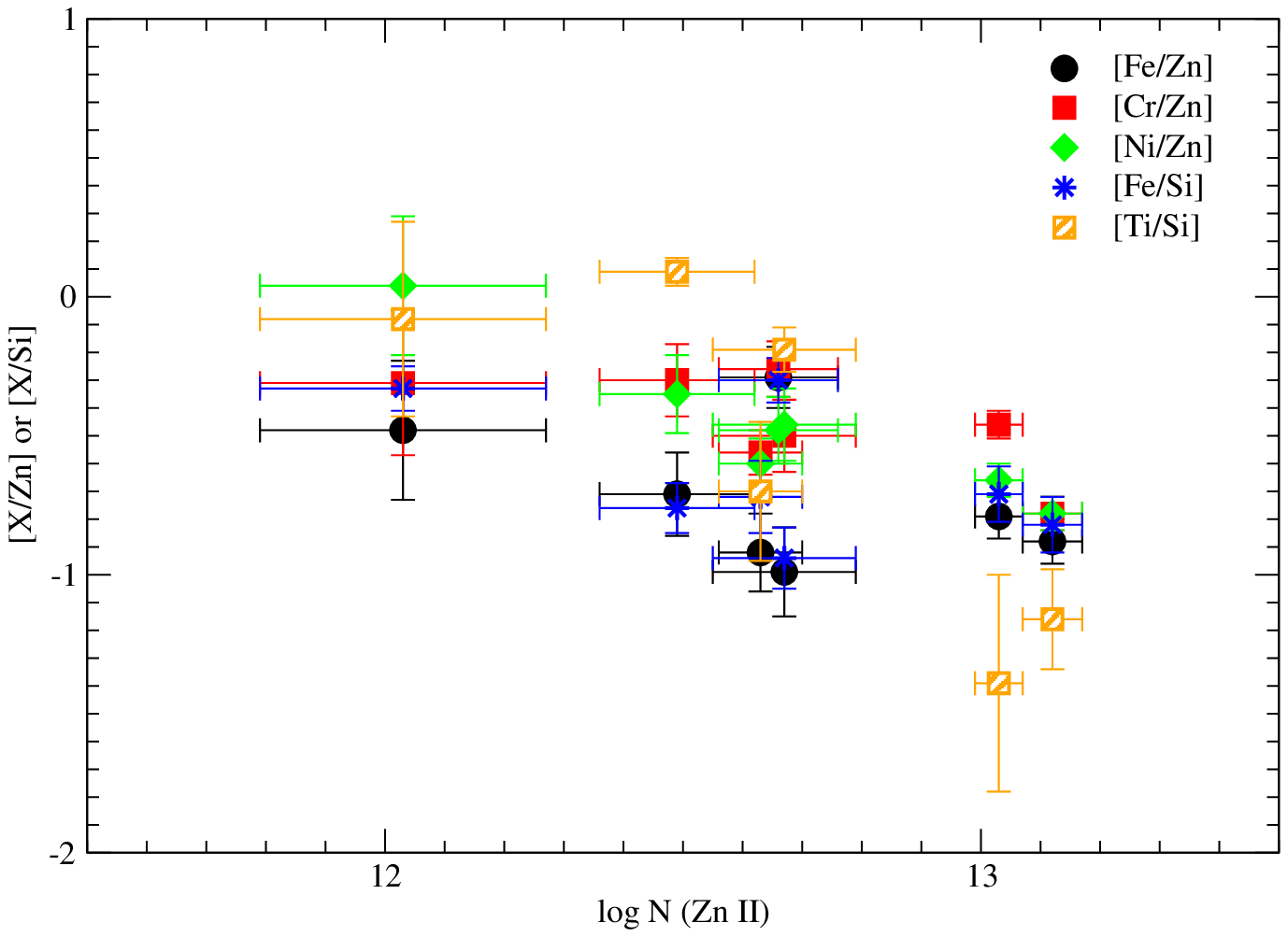}
\hsize=16.0truecm
 \caption{Relative abundances vs. Zn II  column density in different velocity components of the $z_{abs} = 2.2068$ absorber toward Q1135-0010.}
\end{figure}

\newpage
\begin{figure}
\includegraphics[width=0.81\textwidth, angle=270]{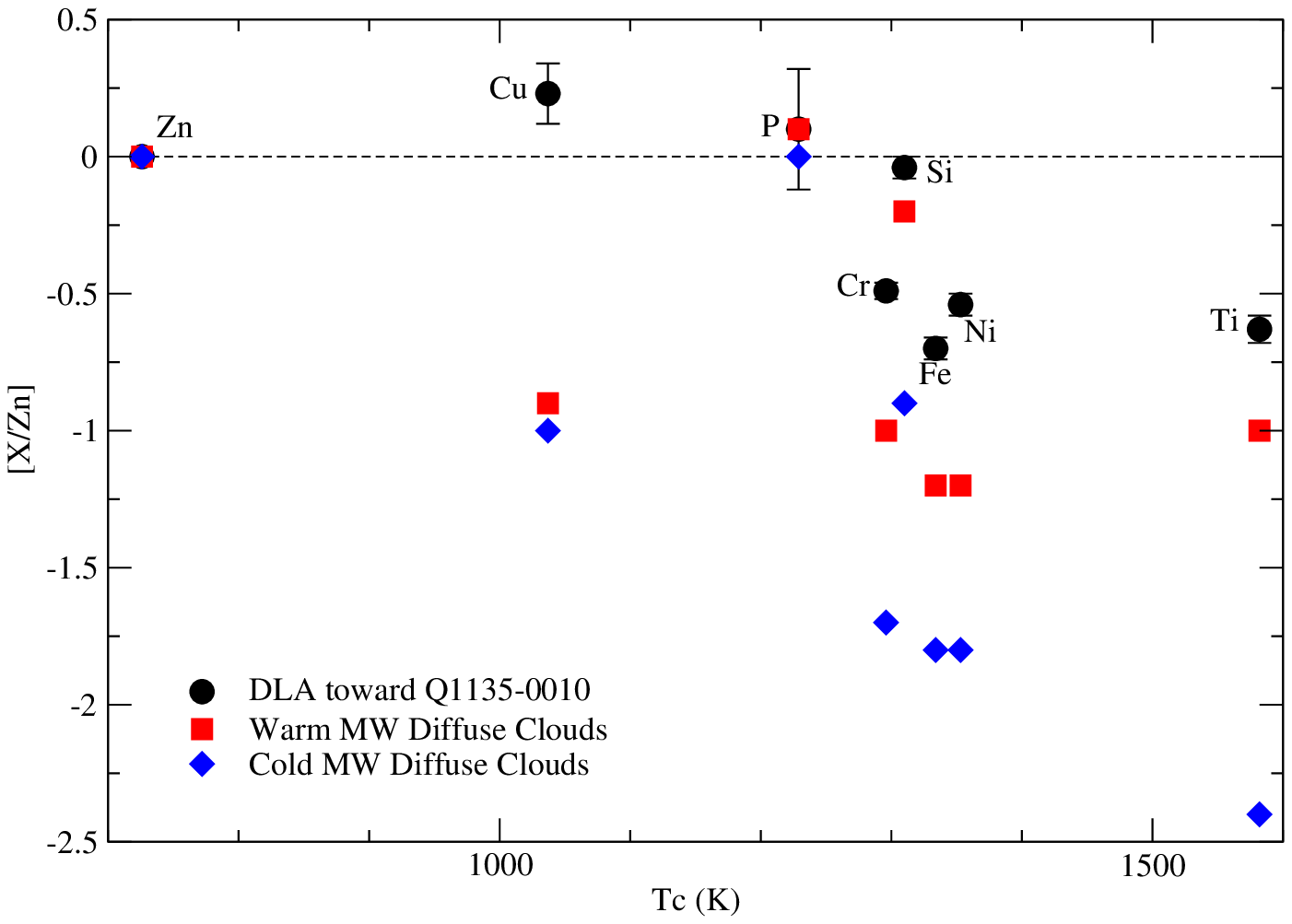}
\hsize=16.0truecm
 \caption{Abundances of various elements relative to Zn, after summing over all the velocity components, plotted vs. condensation temperature. Three points are plotted for each element: the black circles denote the values, along with 1 $\sigma$ uncertainties, for the $z_{abs} = 2.2068$ DLA toward Q1135-0010, while the red squares and blue diamonds denote, respectively, the corresponding relative abundances for warm and cold diffuse interstellar clouds in the Milky Way. }
\end{figure}

\newpage
\begin{figure}
\hskip -0.15in
\includegraphics[width=0.81\textwidth, angle=90]{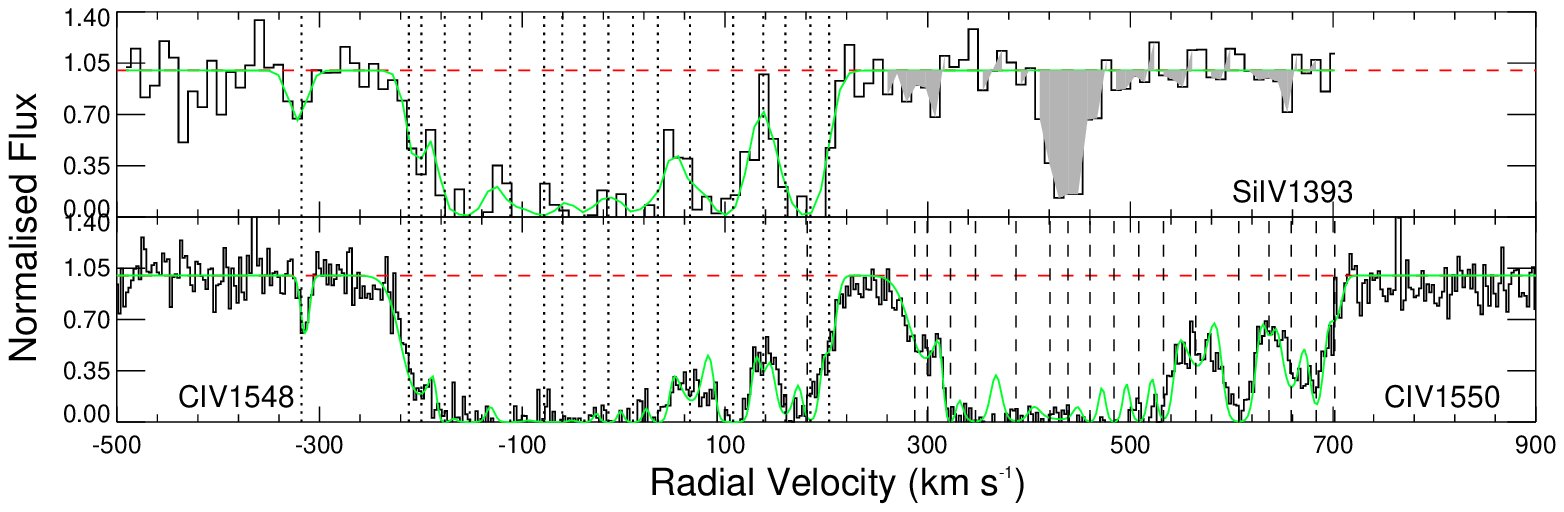}
\vskip -2.0in
 \caption{Velocity plots of higher ionization species in the $z_{abs} = 2.2068$ absorber toward  Q1135-0010. The green curves indicate the multi-component Voigt profile fits. The dotted vertical lines indicate the positions of the velocity components. For C IV, the dashed vertical lines indicate the positions of the velocity components in C IV $\lambda$ 1550.} 
\end{figure}

\newpage
\begin{figure}
\includegraphics[width=0.55\textwidth, angle=270]{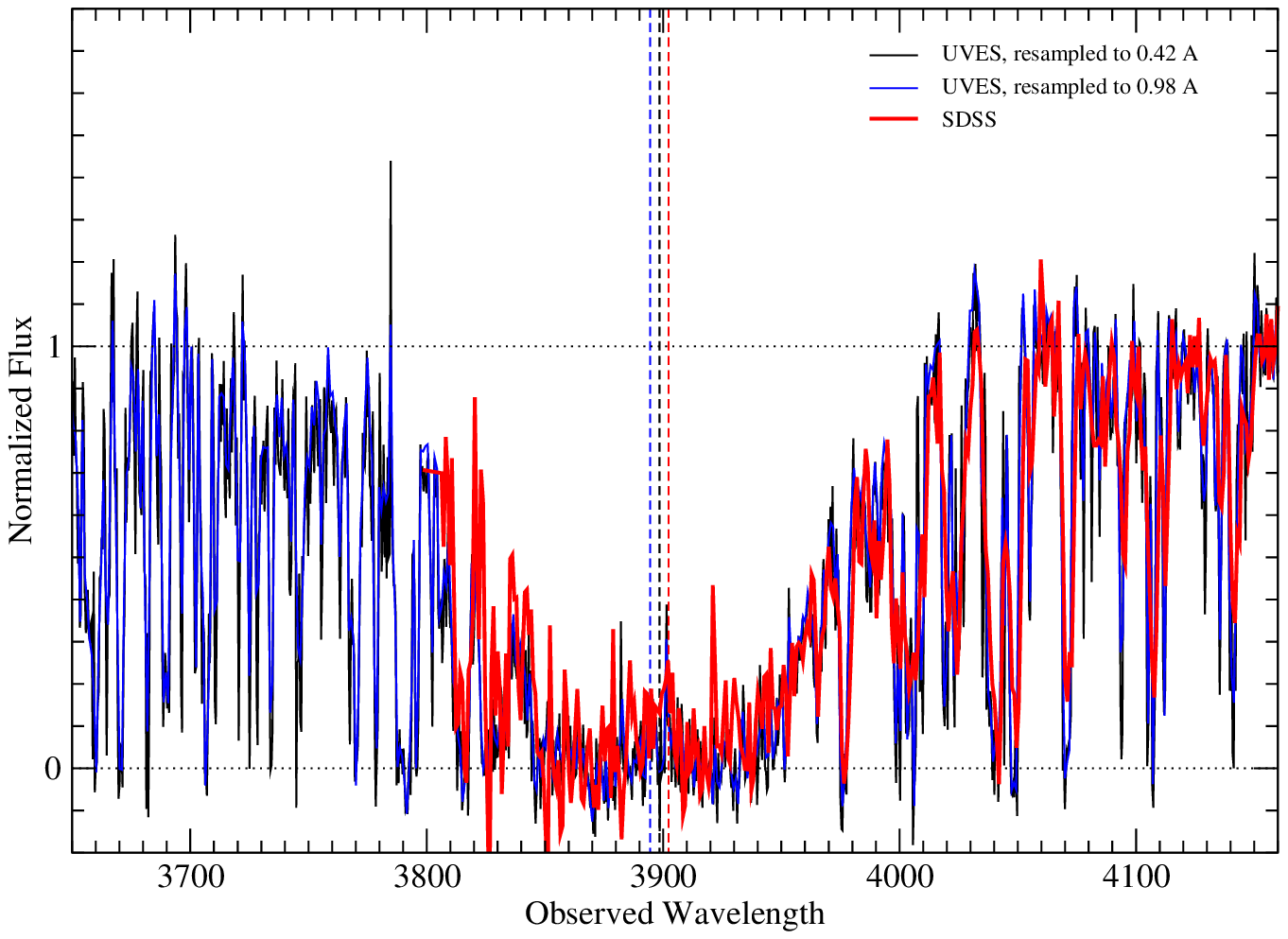}
\vskip -0.3in
\includegraphics[width=0.55\textwidth, angle=270]{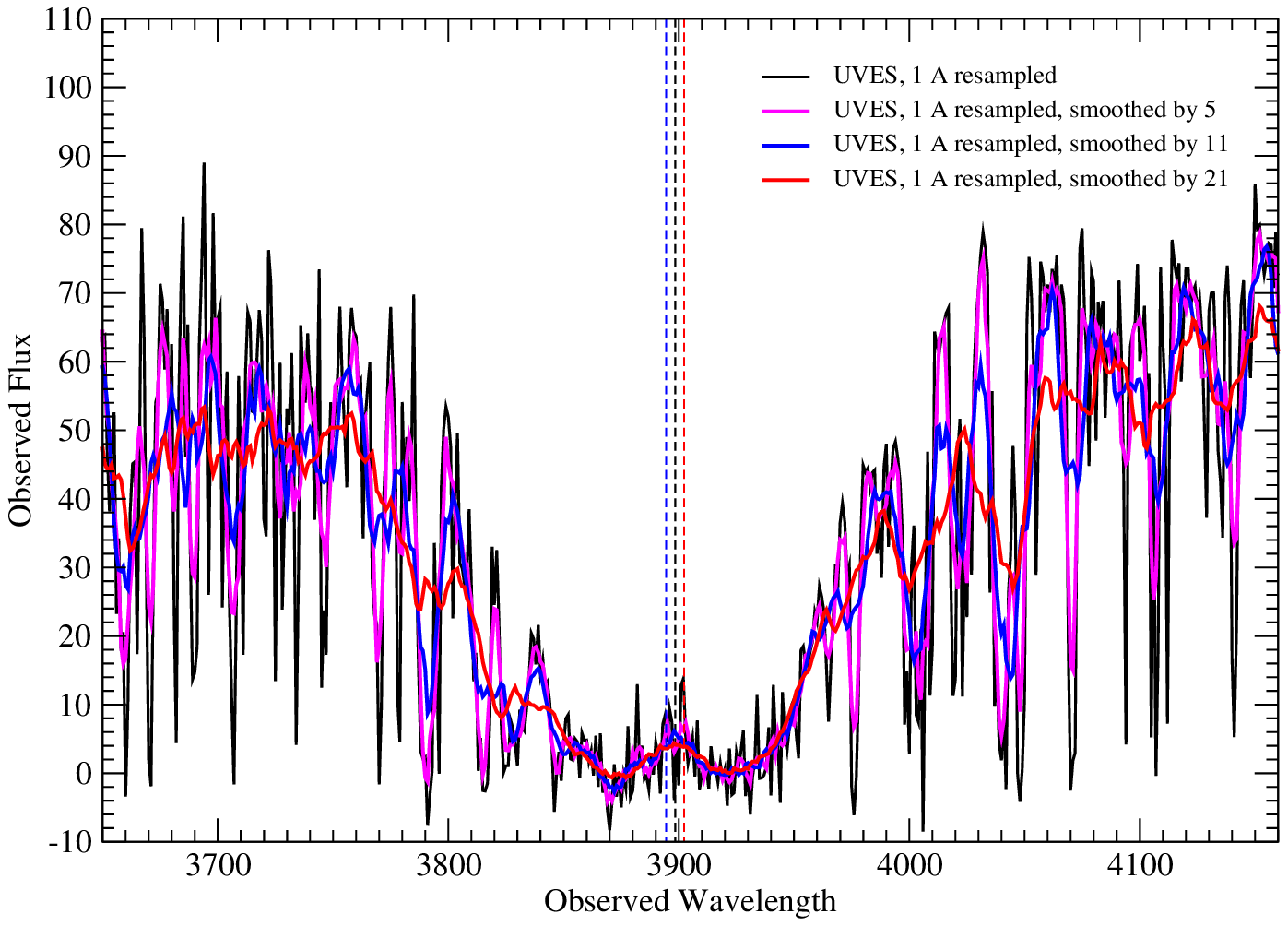}
\hsize=16.0truecm
 \caption{The tentative Lyman-$\alpha$ emission feature seen near the center of the DLA trough.  (a) Top: A closer look at the region near the bottom of the DLA profile. Shown are VLT UVES data resampled to 0.42 and 0.98 {\AA} resolution, with no further smoothing. Also overplotted is the unsmoothed SDSS spectrum. (b) Bottom: VLT UVES data resampled to 1 {\AA} resolution, with and without box-car smoothing. The vertical dashed lines indicate expected position of Ly-$\alpha$ emission at velocity separations of   0, +300, and -300 km s$^{-1}$ with respect to $z=2.2068$.}
\end{figure}

\newpage
\begin{figure}
\includegraphics[width=0.81\textwidth, angle=270]{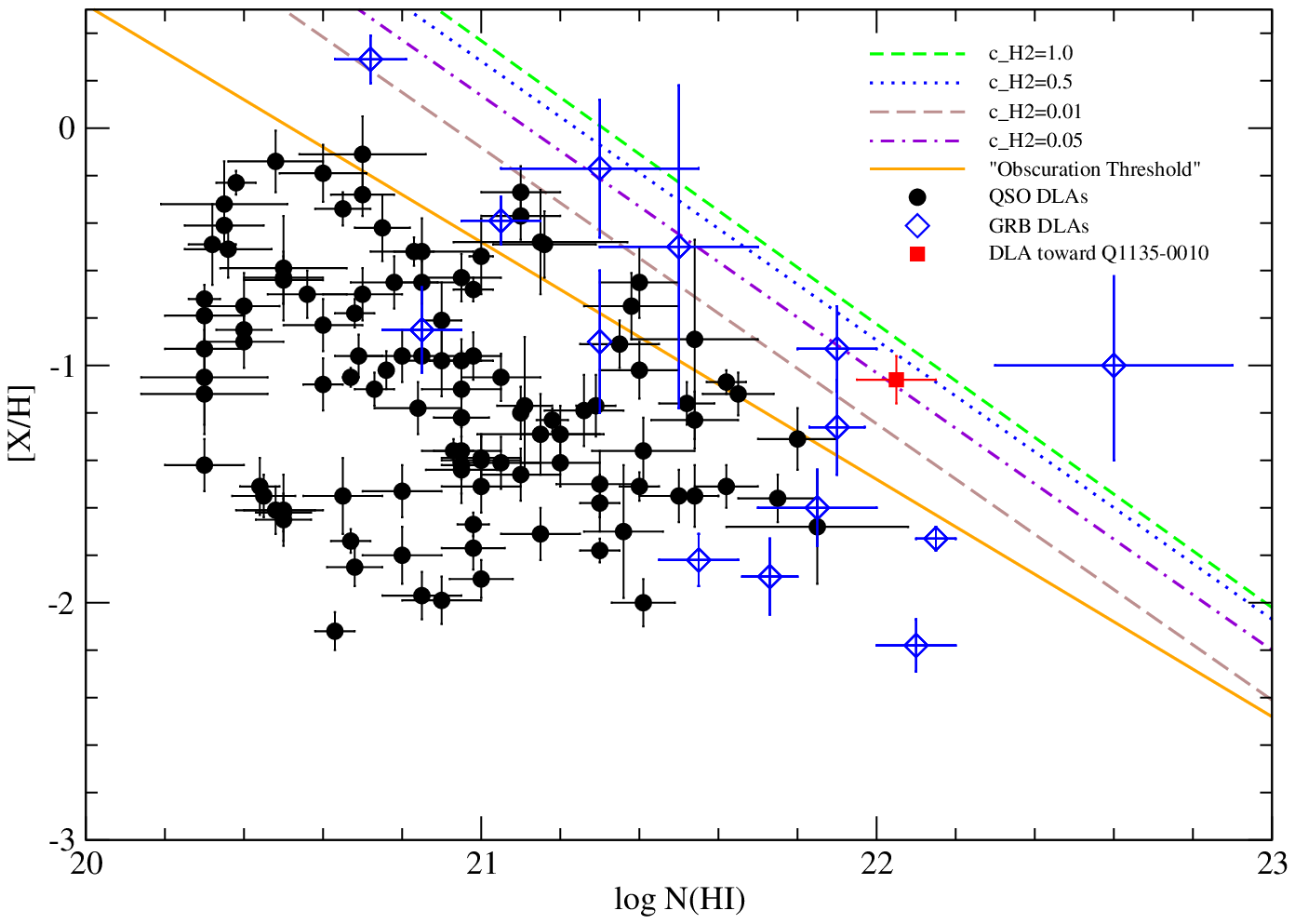}
\hsize=16.0truecm
 \caption{Metallicity (based on firm detections of Zn or S) vs. log $N_{\rm H I}$ for QSO DLAs, GRB DLAs, and the DLA toward Q1135-0010.  The solid orange line shows the ``obscuration threshold'' suggested by Boiss\'e et al. (1998). The remaining lines indicate the curves corresponding to molecular Hydrogen core ``covering fractions" of 0.01, 0.05, 0.5, and 1.0, adopted from Krumholz et al. (2009a). The super-DLA toward Q1135-0010 is clearly well above the ``obscuration threshold", and consistent with $H_{2}$ core ``covering fraction" of $\approx 0.5$.}
\end{figure}

\newpage
\begin{figure}
\includegraphics[width=0.81\textwidth, angle=270]{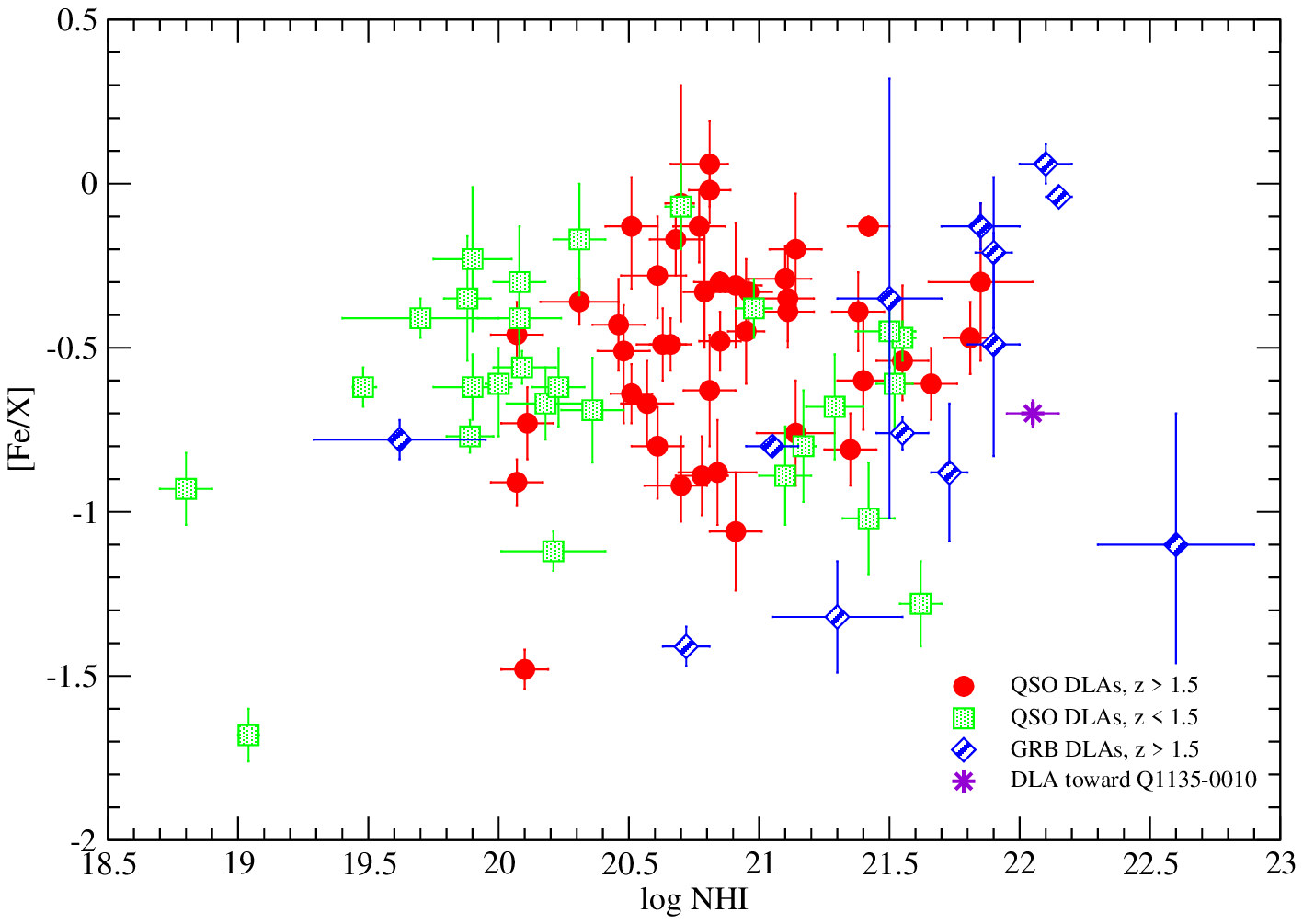}
\hsize=16.0truecm
 \caption{Depletion of Fe relative to Zn or S plotted vs.  H I column density for QSO DLAs at $z < 1.5$, QSO DLAs at  $z >1.5$, GRB DLAs at $z > 1.5$, and the DLA toward Q1135-0010. Error bars denoting 1 $\sigma$ uncertainties are shown where available.}
\end{figure}

\newpage
\begin{figure}
\includegraphics[width=0.81\textwidth, angle=270]{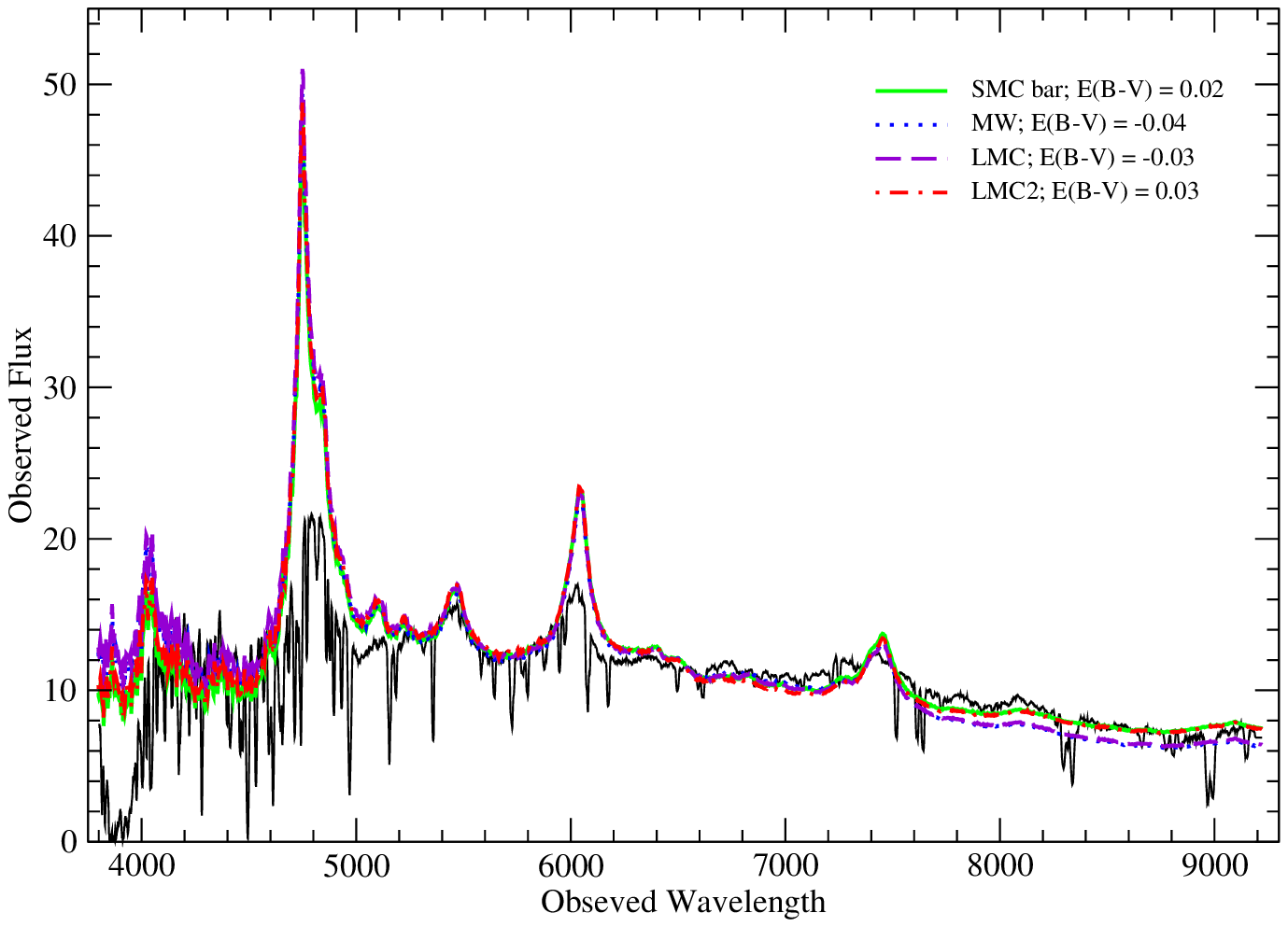}
\hsize=16.0truecm
 \caption{SDSS spectrum of Q1135-0010 after dereddening for Galactic extinction, along with the best-fitting Milky Way, LMC, LMC supershell (LMC2), and SMC bar extinction curves. The SDSS spectra have been box-car smoothed here by a factor of 9 for display purposes. For each extinction curve, the intrinsic QSO spectrum, assumed to be the composite QSO spectrum from Vanden Berk et al. (2001), was reddened as a function of $E(B-V)$ and flux normalization factor. The curves plotted correspond to the best-fitting parameters that give the least squares fit to the data.}
\end{figure}

\newpage
\begin{figure}
\includegraphics[width=0.81\textwidth, angle=270]{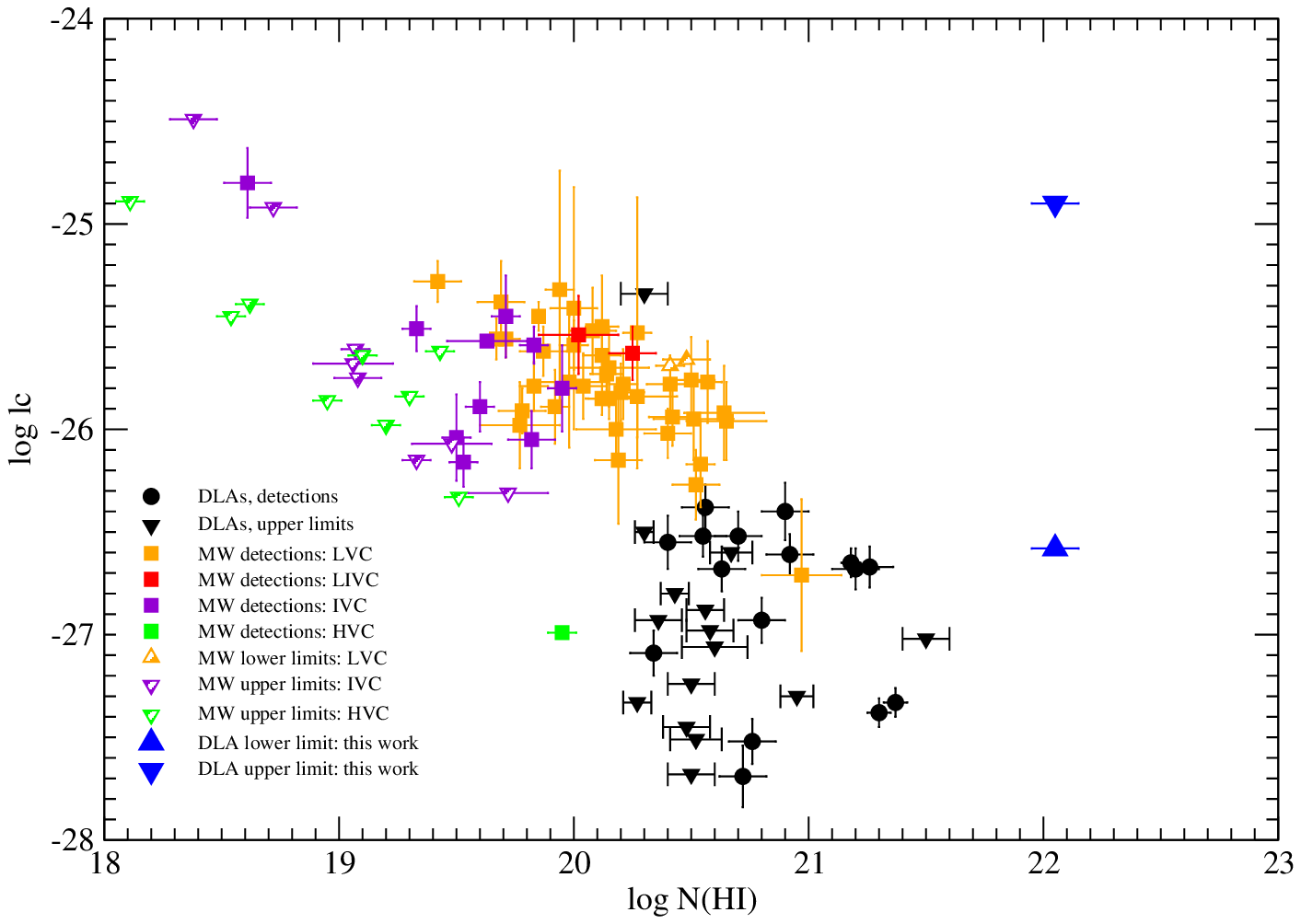}
\hsize=16.0truecm
 \caption{Cooling rate inferred from C II$^{*}$ absorption line plotted vs. H I column density. The larger blue triangles show the estimated limits to the cooling rate we estimate for the DLA toward Q1135-0010. The small black circles and triangles are detections and upper limits, respectively, for the sample of QSO DLAs in Wolfe et al. (2003). The small colored squares denote the measurements for low, intermediate, low+intermediate, and high-velocity interstellar H I clouds in the Milky Way  compiled in Lehner et al. (2004). 
  }
\end{figure}

\newpage
\begin{figure}
\includegraphics[width=0.81\textwidth, angle=90]{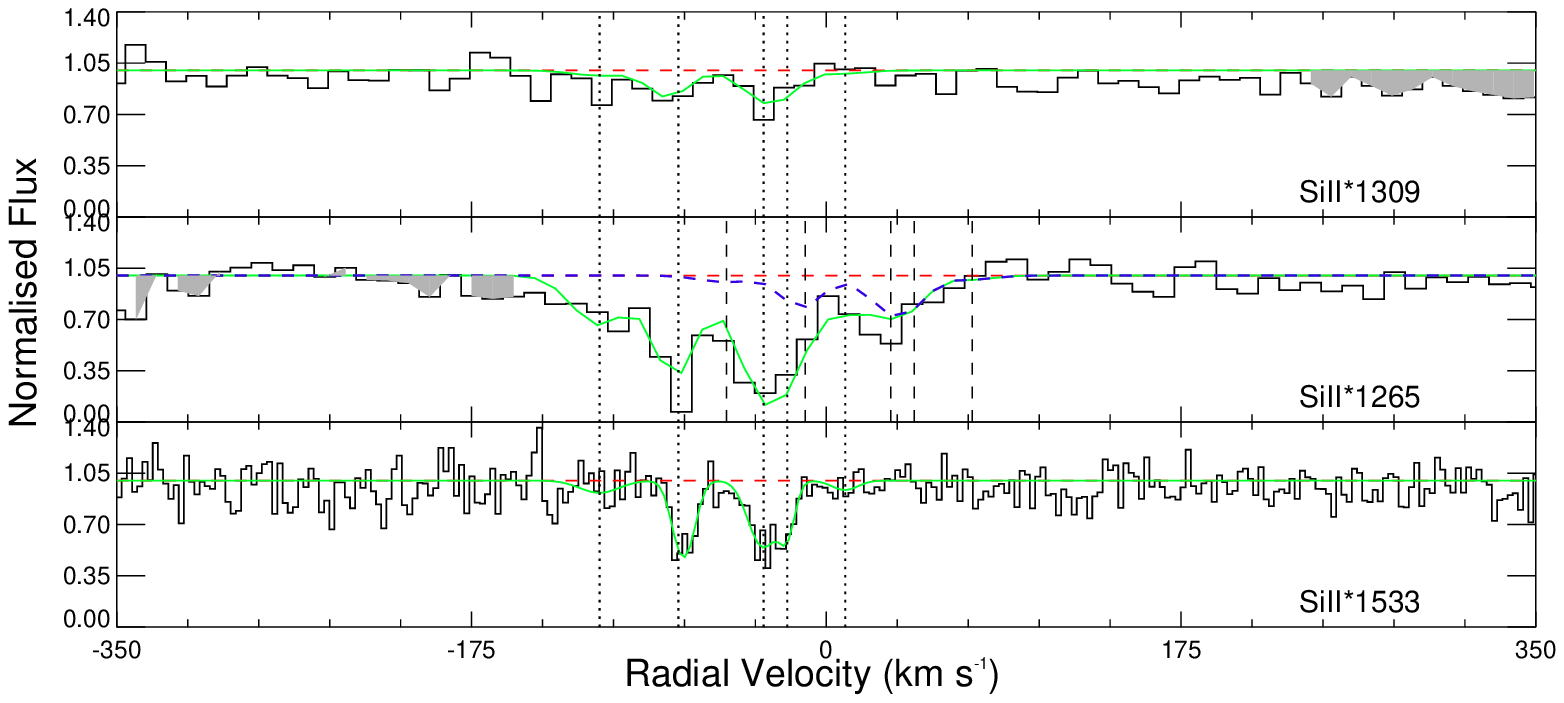}
\vskip -1.0in
 \caption{Velocity plots of Si II$^{*}$ lines in the $z_{abs} = 2.2068$ absorber toward  Q1135-0010. The green curves indicate the multi-component Voigt profile fits. The dotted vertical lines indicate the positions of the velocity components. In the Si II$^{*}$ $\lambda$ 1265 panel, 
 the dashed blue curve denotes the contribution of only 
 Si II$^{*}$ $\lambda$ 1265.0, while the solid green curve denotes the total contribution of both Si II$^{*}$ $\lambda$ 1264.7 and 
 $\lambda$ 1265.0 lines.} 
\end{figure}

{\begin{landscape}
\begin{table}
\caption{Voigt Profile Fitting Parameters for Various Velocity Components for Selected Ions in the $z=2.2068$ DLA}

\begin{tabular}{cccccccc}
\hline\hline
mean vel.	&$b$&	Cr II	&	Zn II	&	Fe II	&	Ni II	&	Si II	\\
(km s$^{-1}$)& (km s$^{-1}$)& (cm$^{-2}$)&(cm$^{-2}$)&(cm$^{-2}$)&(cm$^{-2}$)&(cm$^{-2}$)\\
\hline
-138	& 9.4 &	--	&	--	&	(1.35$\pm$0.17)e14	&	--	&	(2.84$\pm$0.64)e14	 \\
-115.5&13.1	&	(1.60$\pm$0.14)e13	&	(3.07$\pm$0.86)e12	&	(4.17$\pm$0.82)e14	&	(5.31$\pm$0.64)e13	&	(2.79$\pm$0.20)e15 	\\
-75.5	&17.0&	(1.57$\pm$0.19)e13	&	(4.73$\pm$1.23)e12	&	(3.37$\pm$0.81)e14	&	(6.36$\pm$0.87)e13	&	(3.43$\pm$0.31)e15	 \\
-60.0	&11.8& 	(2.60$\pm$0.21)e13	&	(4.57$\pm$1.00)e12	&	(1.63$\pm$0.23)e15	&	(5.86$\pm$0.84)e13	&	(3.78$\pm$0.44)e15	 \\
-39.0	&8.0&	(3.42$\pm$0.98)e12	&	--	&	(2.65$\pm$0.90)e14	&	(1.43$\pm$0.44)e13	&	(7.91$\pm$1.17)e14	\\
-16.5	&7.4&	(2.33$\pm$0.17)e13	&	(1.33$\pm$0.14)e13	&	(1.20$\pm$0.19)e15	&	(8.65$\pm$0.90)e13	&	(9.36$\pm$1.50)e15\\
5.0	&8.3&	(1.22$\pm$0.12)e13	&	(4.22$\pm$0.66)e12	&	(3.48$\pm$0.96)e14	&	(4.13$\pm$0.56)e13	&	(2.16$\pm$0.21)e15\\
30.5	&8.0&	(3.93$\pm$0.25)e13	&	(1.08$\pm$0.10)e13	&	(1.20$\pm$0.18)e15	&	(9.17$\pm$0.91)e13	&	(7.25$\pm$1.15)e15	\\
62.5	&11.2&	(5.44$\pm$1.07)e12	&	(1.07$\pm$0.54)e12	&	(2.44$\pm$0.26)e14	&	(4.55$\pm$0.58)e13	&	(6.12$\pm$0.91)e14 \\
103.0&12.4&	--	&	--	&	(1.58$\pm$0.43)e13	&	(1.06$\pm$0.46)e13	&	(5.50$\pm$0.47)e13	 \\
145.0&11.3&	--	&	--	&	--	&	--	&	(2.33$\pm$1.71)e12\\
227.5&15.5	&	--	&	--	&	(2.24$\pm$0.53)e13	&	--	&	(3.97$\pm$0.38)e13\\

\hline
\hline
\end{tabular}

\end{table}
\end{landscape}

{\begin{landscape}
\begin{table}
\caption{Voigt Profile Fitting Parameters for Various Velocity Components for Additional Ions in the $z=2.2068$ DLA}

\begin{tabular}{ccccccccc}
\hline\hline
mean vel.	& $b$&C II$^{\dagger}$ & C II$^{*}$& O I$^{\dagger}$ &Ti II& P II & Cu II\\
(km s$^{-1}$)& (km s$^{-1}$)& (cm$^{-2}$)&(cm$^{-2}$)&(cm$^{-2}$)&(cm$^{-2}$)&(cm$^{-2}$)&(cm$^{-2}$)\\
\hline

-138	& 9.4	& --& --& $>$ 3.45e14& --& --& -- \\
-115.5&13.1	 & 8.68e18&5.57e14&$>$ 1.39e15&(8.3 $\pm$ 0.72)e12&(1.74 $\pm$ 1.08)e13&--	\\
-75.5 &17.9	&7.56e18& 1.84e14&$>$ 8.82e14&(5.28 $\pm$ 0.88)e12&(3.48 $\pm$ 3.09)e13& (4.97 $\pm$3.37)e12\\
-60.0&11.8 	&2.32e18 & --&$>$ 2.58e15&---&(1.22 $\pm$ 1.11) e14 &(5.22 $\pm$  31.20)e11\\
-39.0&8.0 	& --& --& $>$ 6.55e14&(9.69 $\pm$4.85)e11&(5.63$\pm$6.77)e13&(1.33 $\pm$ 0.37)e13\\
-16.5&7.4     & 3.55e18 &--&$>$ 8.93e15&(1.54 $\pm$0.56)e12&(9.37 $\pm$9.82)e13&	(4.01 $\pm$ 2.54)e12\\
5.0&8.3	        & --& 1.04e14&$>$ 7.88e14&(1.04 $\pm$0.54)e12&--&(1.65 $\pm$ 2.07)e12 \\
30.5&8.0  	&6.72e17& 1.05e14&$>$ 4.45e15&(7.02 $\pm$4.97)e11&(3.57 $\pm$ 7.81)e12& (4.88 $\pm$ 17.40)e11\\
62.5&11.2	&-- & 2.23e13	& $>$ 5.88e13&(1.22 $\pm$0.80)e12&--&--\\
103.0&12.4	&5.00e14&5.20e12&$>$ 2.59e12&--&--&(6.84 $\pm$ 18.30)e11 \\
145.0&11.3	&1.06e15	&9.58e12&--&--&--&(6.81 $\pm$17.90)e11\\
227.5&15.5	& 5.30e16&1.19e13&--&--&(2.81 $\pm$ 1.40)e13&(3.56 $\pm$2.09)e12	\\

\hline
\hline
\end{tabular}

$\dagger$ The column densities for C II, C II*, and O I in some components could be higher because of line saturation. 
\end{table}
\end{landscape}

\begin{table} 
\caption{Total Gas-phase Column Densities and Abundances in  the $z=2.2068$ DLA}
\begin{tabular}{ccccc}
\hline\hline
Element &log (X/H)$_{\odot}$& log N$_{total}$ & [X/H] &[X/Zn]\\
\hline
C&-3.61&$< 19.36$&$< 0.92$&$< 1.98$\\
O&-3.31&$> 16.30$&$> -2.44$&$> -1.38$\\
Si&-4.46&$16.49 \pm 0.03$&$-1.10 \pm 0.10$&$-0.04 \pm 0.04$\\
P&-6.54&$14.58 \pm 0.21$&$-0.93 \pm 0.23$&$0.13 \pm 0.21$\\
Ti&-7.08&$13.28 \pm  0.04$&$-1.69 \pm 0.11$&$-0.63 \pm 0.05$\\
Cr&-6.35&$14.15 \pm 0.02$&$-1.55 \pm 0.10$&$-0.49 \pm 0.03$\\
Fe&-4.53&$15.76 \pm 0.03$&$-1.76 \pm 0.10$&$-0.70 \pm 0.04$\\
Ni&-5.78&$14.67 \pm 0.02$&$-1.60 \pm 0.10$&$-0.54 \pm 0.03$\\
Cu&-7.74&$13.56 \pm 0.10$&$-0.75 \pm 0.14$&$+0.31 \pm 0.10$\\
Zn&-7.37&$13.62 \pm 0.03$&$-1.06  \pm 0.10$&0.0\\

\hline
\hline
\end{tabular}
\end{table}

\begin{table}
\caption{Voigt Profile Fitting Parameters for C IV, Si IV in the $z=2.2068$ DLA$\dagger$}

\begin{tabular}{ccccc}
\hline\hline
mean vel.	& $b_{\rm Si IV}$& $N_{\rm Si IV}$ & $b_{\rm C IV}$& $N_{\rm C IV}$ \\
(km s$^{-1}$)& (km s$^{-1}$)& (cm$^{-2}$) &(km s$^{-1}$)& (cm$^{-2}$) \\
\hline
-318&4.6&(6.16 $\pm 3.39$)e12&4.6&(6.55 $\pm 1.53$)e12\\
-212 &... & ... &19.6 & $>$1.51e13\\
-202&4.9 & $>$3.05e13& 4.9&$>$5.55e13\\
-176 &4.8 &$>$8.44e13&4.8&$>$ 1.78e14\\
-152 &13.3&$>$1.07e14 &13.3&$>$3.38e14\\
-112 &13.1 &$>$3.51e13 & 13.1&$>$2.94e14\\
-78& 17.0 &$>$9.79e13 & 17.0 & $>$2.58e14\\
-60 &11.8 & $>$2.93e12& 11.8 & $>$1.35e14\\
-39 & 8.0 & $>$1.26e14 & 8.0 & $>$2.15e14\\
-15 & 7.4 & $>$2.07e13 & 7.4 & $>$2.35e14\\
9 & 8.3 & $>$1.18e14 & 8.3 & $>$2.09e14\\
34 & 8.0 & $>$2.53e13 & 8.0 & $>$1.49e14\\
60 &...&...&17.7 & $>$5.12e13\\
68 & 11.2 & $>$2.03e13 & 11.2 & $>$2.48e13\\
105 & 15.3 & $>$1.07e14&12.4& $>$3.26e14\\
138&...&...&6.3&$>$1.48e13\\
157& 11.3 & $>$9.76e12& 11.3 & $>$7.68e13\\
181 & 13.3 &$>$ 1.31e14& 13.3&$>$7.96e13\\
203&...&...&7.6&$>$1.32e13\\

\hline
\end{tabular}

$\dagger$ The column densities for C IV and Si IV in most components could be higher because of line saturation. 

\end{table}

\end{document}